\documentclass[12pt,a4paper,oneside]{article}
\usepackage{graphicx,amssymb,amsmath,color}
\usepackage[linkcolor={blue},citecolor={red},colorlinks=true]{hyperref}
\usepackage{cite}
\usepackage{relsize}
\usepackage{enumerate}
\usepackage[margin=1.5 cm]{geometry}
\usepackage{slashed}
\usepackage{multirow}
\usepackage{subfigure}
\usepackage[normalem]{ulem}
\usepackage[dvipsnames, x11names]{xcolor}
\usepackage[compat=1.0.0]{tikz-feynman}
\usepackage{xcolor}
\usepackage{mathtools}

\def\beq{\begin{equation}}
\def\eeq{\end{equation}}
\def\bea{\begin{eqnarray}}
\def\eea{\end{eqnarray}}

\def\bit{\begin{itemize}}
\def\eit{\end{itemize}}

\def\baa{\begin{array}}
\def\eaa{\end{array}}

\def\simgt{\mathrel{\lower2.5pt\vbox{\lineskip=0pt\baselineskip=0pt
           \hbox{$>$}\hbox{$\sim$}}}}
\def\simlt{\mathrel{\lower2.5pt\vbox{\lineskip=0pt\baselineskip=0pt
           \hbox{$<$}\hbox{$\sim$}}}}

\def\bfc{\begin{figure}\begin{center}}
\def\efc{\end{center}\end{figure}}
\def\nn{\nonumber\\}

\definecolor{chromeyellow}{rgb}{1.0, 0.65, 0.0}
\definecolor{darkcoral}{rgb}{0.8, 0.36, 0.27}
\definecolor{cadmiumgreen}{rgb}{0.0, 0.42, 0.24}

\begin{document}
\leftline{DESY-24-095}

\begin{flushright}
\hspace{3cm} 
\end{flushright}
\vspace{.6cm}
\begin{center}

\hspace{-0.4cm}{\LARGE \bf 
ALP leptogenesis\\
\vspace{.2 cm}
\large
non-thermal right-handed neutrinos from axions
}

\vspace{1cm}{Martina Cataldi$^{a,b,1}$, Alberto Mariotti$^{c,2}$, Filippo Sala$^{d,e,3}$, Miguel Vanvlasselaer$^{c, 4}$}
\\[7mm]

{$^a$ \it Deutsches Elektronen-Synchrotron DESY, Notkestr.\,85, 22607 Hamburg, Germany}\\

{$^b$ \it II. Institute of Theoretical Physics, Universität Hamburg, Luruper Chaussee 149, 22761, Hamburg, Germany}\\

{$^c$ \it Theoretische Natuurkunde and IIHE/ELEM, Vrije Universiteit Brussel,
\& The International Solvay Institutes, Pleinlaan 2, B-1050 Brussels, Belgium }

{$^d$ \it
Dipartimento di Fisica e Astronomia, Universit\`a di Bologna, via Irnerio 46, 40126 Bologna, Italy}\\

{$^e$ \it
INFN, Sezione di Bologna, viale Berti Pichat 6/2, 40127 Bologna, Italy
}

\end{center}

\bigskip \bigskip \bigskip

\centerline{\bf Abstract} 
\begin{quote}
We propose a novel realisation of leptogenesis that relies on the out-of-equilibrium decay of an axion-like particle (ALP) into right-handed Majorana neutrinos (RHNs) in the early Universe.
With respect to standard thermal leptogenesis, our mechanism lowers by two orders of magnitude the RHN mass, or the tuning in the RHN mass splittings, needed to reproduce the baryon asymmetry of the Universe and neutrino masses.
We find that ALP leptogenesis requires $m_a > 10^{4}$ GeV and $f_a > 10^{11}$ GeV  for the ALP mass and decay constant, and predicts an early period of matter domination induced by the ALP in parts of its parameter space.
We finally provide a viable supersymmetric realisation of ALP leptogenesis where the ALP is the $R$-axion, which accommodates GeV gravitino dark matter and predicts RHN below 10 TeV.\\

\end{quote}

\vfill
\noindent\line(1,0){188}
{\scriptsize{ \\ E-mail:
\texttt{$^1$\href{}{martina.cataldi@desy.de}},
\texttt{$^2$\href{}{alberto.mariotti@vub.be}},
\texttt{$^3$\href{}{f.sala@unibo.it}},
\texttt{$^4$\href{}{miguel.vanvlasselaer@vub.be}}}
}

\newpage

\tableofcontents

\section{Introduction}

The baryon asymmetry of the universe (BAU) remains one of the unsolved puzzles of our understanding of Nature. Its numerical value, obtained via the latest measurements of the cosmic microwave background (CMB)~\cite{Planck:2015fie} and of the primordial abundance of light elements~\cite{Fields:2019pfx}, reads
\bea 
\label{eq:observed_BAU}
Y_{\Delta B} \equiv \frac{n_{B}-n_{\overline{B}}}{s} \bigg|_{0} \approx (8.69 \pm 0.22) \times 10^{-11}
\eea 
where $n_{B}$, $n_{\overline{B}}$ and $s$ are the number densities of baryons, antibaryons and entropy evaluated at present time.

Setting the BAU as an initial condition in the very early universe is incompatible with the inflationary paradigm, which would dilute it away. One then needs to generate it via a so-called `baryogenesis' mechanism occurring after inflation, which needs to satisfy the Sakharov requirements~\cite{Sakharov:1967dj}:  i) violation of Standard Model (SM) baryon number, ii) violation of $C$
and $CP$, and iii) departure from equilibrium in the early universe. 
In principle, the SM fulfils all the aforementioned conditions: the electroweak (EW) or QCD phase transitions (PTs) might bring out-of-equilibrium effects, the EW sector is chiral and maximally breaks $C$, the CKM matrix contains $CP$ violation and sphalerons violate the conservation of baryon number.
However, within the SM, departure from equilibrium is too mild both at the QCD~\cite{Aoki:2006we} and the EW~\cite{Kajantie:1996mn} PTs and the $CP$ violation in the CKM matrix too suppressed to account for the observed BAU~\cite{Gavela:1993ts}. 
As a consequence the BAU requires physics beyond the SM (BSM).

Among the many proposed BSM mechanisms for baryogenesis (see e.g.~\cite{Riotto:1998bt,Buchmuller:2021} for reviews), leptogenesis~\cite{FUKUGITA198645} explains at once the BAU and neutrino oscillations, another major shortcoming of the SM.
It does so via new fermions, singlets under the SM gauge group and dubbed `Right-handed neutrinos' (RHN), whose early-universe decays source a lepton asymmetry which is later partially converted into a baryon asymmetry through SM sphaleron processes, see~\cite{Davidson:2008bu, Buchm_ller_2005} for reviews.

In the same spirit of addressing more SM shortcomings within a common picture, it is natural to go further and ask how leptogenesis may change when it is embedded in BSM frameworks that are motivated independently of the BAU or of neutrino oscillations. 
In this paper we answer such question in the case a BSM axion-like particle (ALPs) exists in the spectrum and is sizeably produced in the early universe. ALPs, or pseudo-Goldstone bosons, are a generic prediction of any BSM theory where a global symmetry is broken spontaneously at some scale $f_a$, and at the same time mildly broken explicitly so they get a mass $m_a \ll 4\pi f_a$. Here we will be interested in the case where ALPs have mass and couplings that allow them to sizeably decay to RHNs in the early universe, and so generate a non-thermal RHN population that alters the dynamics of standard thermal leptogenesis\footnote{Leptogenesis from a non-thermal population of RHN can also be realised, for example, from inflaton decays~\cite{Chung:1998rq,Giudice:1999fb,Hahn-Woernle:2008tsk, Zhang:2023oyo}, bubble wall dynamics at first-order PTs~\cite{Katz:2016adq,Azatov:2021irb, Huang:2022vkf, Chun:2023ezg, Dichtl:2023xqd}, or axion domain-wall dynamics \cite{Daido:2015gqa}.}. We anticipate that this will restrict ourselves to ALPs with a mass above the TeV scale\footnote{Much lighter ALPs can affect leptogenesis via their background field behaviour, see e.g. the recent~\cite{Chun:2023eqc}.}.
Such `heavy' ALPs are predicted for a variety of reasons which are independent of the BAU and neutrino oscillations, for example (see e.g.~\cite{CidVidal:2018blh} for a synthetic review): i) to address the strong $CP$ problem with an axion~\cite{Weinberg:1977ma,Wilczek:1977pj} and at the same time solve its axion quality problem~\cite{Kamionkowski:1992mf,Holman:1992us,Ghigna:1992iv,Barr:1992qq}; ii) as an $R$-axion, which is a generic prediction of low-energy supersymmetry (SUSY) breaking~\cite{Nelson:1993nf,Intriligator:2007py,Bellazzini:2017neg}, SUSY being motivated by unification of matter and interactions, unification of gauge couplings, the big hierarchy problem, and as a necessary ingredient for string theory, see~\cite{Martin:1997ns,Terning:2006bq} for reviews.
Despite heavy ALPs received a lot of attention in recent literature, to our knowledge their connection with leptogenesis has not been explored so far.

Key new findings of our `ALP leptogenesis' mechanism include that the RHN mass needed to realise the BAU is lowered with respect to standard thermal leptogenesis, that an ALP-induced period of early matter domination is predicted in part of the viable parameter space, and that the SUSY embedding predicts a very sharp connection with dark matter in the form of gravitinos with GeV mass.

\begin{figure}[h!]
\centering
\includegraphics[width=17cm]{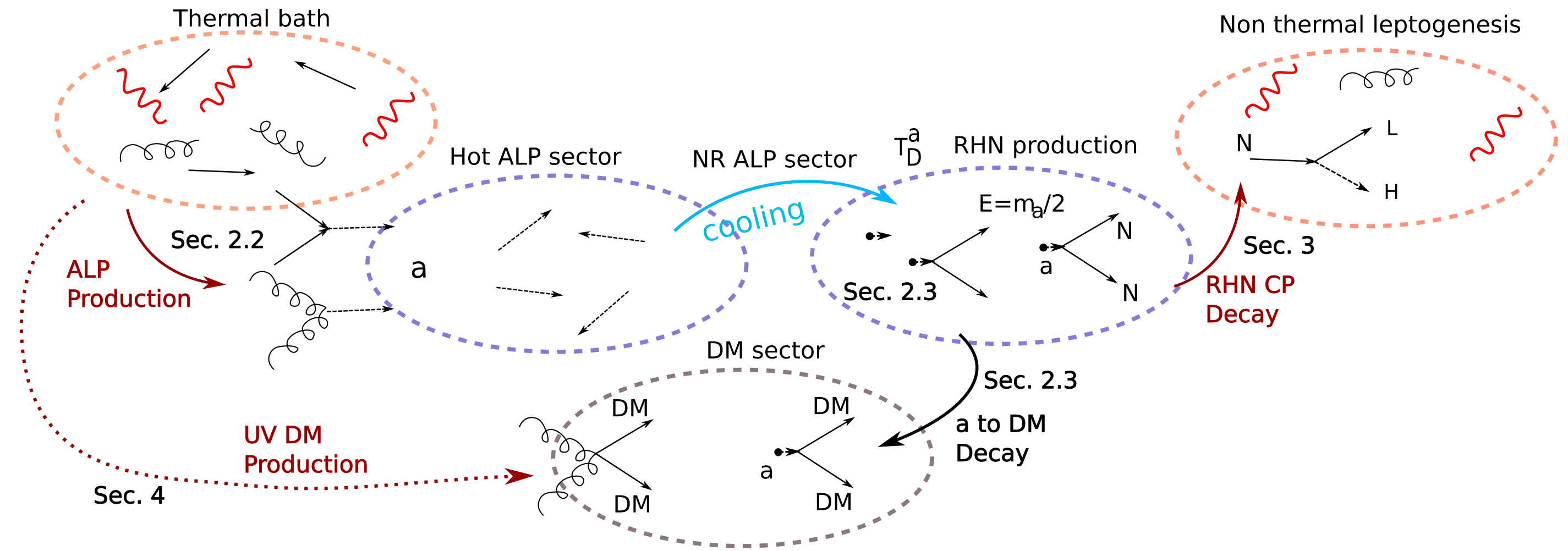}

\caption{\small{Sketch of ALP leptogenesis. The ALP is produced by the thermal sector, and then dominantly decays to non-thermal RHNs (and to DM in the SUSY embedding). The RHNs decay back to the thermal sector while violating $CP$ and inducing leptogenesis. The section in which each of the processes described are discussed is also mentioned on the figure.
}}
 \label{fig:schematic}
\end{figure}

 This article is organized as follows: in Sec.~\ref{sec:Alp_prod} we introduce ALPs and discuss their production and decays in the early-universe, in
  Sec.~\ref{sec:leptogenesis} we first review standard thermal leptogenesis and then quantitatively present our new mechanism  of ALP leptogenesis, in Sec.~\ref{sec:gravitino} we discuss the further features gained when the ALP in question is the SUSY `R-axion', in Sec.~\ref{sec:summary} we conclude. For easiness of the reader, in Fig.~\ref{fig:schematic} we sketch the early-universe dynamics of ALP leptogenesis, together with the section where each step is discussed.

\section{Thermal history of ALP in the Early Universe}

\label{sec:Alp_prod}

In this section we describe the ALP model that we consider and sketch its thermal history.

\subsection{The ALP model}

We first introduce the main characteristics of the ALP we will consider. In its original realisation, the axion has been introduced to solve the strong $CP$ problem of QCD via the Peccei-Quinn (PQ) mechanism \cite{Peccei:1977hh,Peccei:1977ur,Weinberg:1977ma,Wilczek:1977pj,Zhitnitsky:1980tq,Dine:1981rt}. It has been however soon realised that ALPs, i.e. pseudo-nambu Goldstone bosons of spontaneously broken global symmetries, can emerge in many scenarios of BSM physics. 
ALPs are described by an effective theory expanded in inverse powers of the axion decay constant $f_a$ that fully describes their dynamics and interactions with the SM up to temperatures of order $f_a$.

Here we consider an ALP coupled both to the strong sector of the SM as well as to the RHN. The effective Lagrangian describing the ALP's couplings at first order in the ALP field $a$ reads
\begin{equation}
\label{eq:axionl}
\begin{split}
    \mathcal{L}_a= \, & \frac{1}{2} \partial_{\mu} a \partial^{\mu} a - \frac{1}{2} m_a^2 a^2 - \frac{\alpha_s}{8 \pi} C_{g} \frac{a}{f_a} G_{\mu \nu}^{b} \tilde{G}^{b, \mu \nu} +   \frac{\partial_{\mu}{a}}{f_{a}} C_{t}\overline{t}_R \gamma^{\mu}  t_R +\\
     & + \frac{\partial_{\mu}{a}}{f_{a}} C_{Q_3}\overline{Q}_3 \gamma^{\mu}  Q_3 
     + 
    \frac{\partial_{\mu}a}{f_{a}} \overline{N}_R \gamma^{\mu}  N_R
\end{split}
\end{equation}
where $G_{\mu \nu}^b$ is the gluon field strength with its dual $\tilde{G}_{\mu \nu} = \frac{1}{2} \epsilon_{\mu \nu \alpha \beta} G_{\alpha \beta}$, $Q_3$ is the left-handed doublet of the third quark generation, $\alpha_s=g_s^2 / (4 \pi)$, $t_R$ and $N_R$ are the right-handed top quark and neutrino. The coupling constants $C_g$, $C_{Q_3}$ and $C_t$ are dimensionless coefficients. The last term in Eq.\eqref{eq:axionl} denotes the interaction between the ALP and the RHN,
where for simplicity we have set to $1$ the dimensionless coupling.

The ALP mass and the decay constant will first be treated as free independent parameters, but always restricting to the range
$
m_a > 2 M_N
$, such that the decay into the RHN is kinematically allowed.

\paragraph{Within SUSY realisations}

A concrete realization of the ALP described above is given by the R-axion, the pseudo nambu-goldstone boson related to the breaking of the R-symmetry in SUSY models.
This will be presented in detail in Sec.~\ref{sec:gravitino}, for clarity we anticipate some basic elements here.
As observed in \cite{Nelson:1993nf,Intriligator:2007py}, 
SUSY breaking sectors naturally deliver in the low energy spectrum an R-axion, the PNGB of the spontaneously broken R-symmetry, whose effective action can be fully captured in terms of constrained superfields. The R-axion mass receives an irreducible contribution from the tuning of the cosmological constant which is proportional to the SUSY breaking scale and is hence related to the gravitino mass,
which reads
\cite{Bellazzini:2017neg}
\bea 
\label{eq:irred}
(m^{\rm irred}_a)^2 \approx 
\frac{24 m_{3/2} \omega_R}{f_a^2} \, ,
\qquad
\omega_R < \frac{f_a F}{2 \sqrt{2}} \, ,
\eea 
where $m_{3/2} = F/\sqrt{3} M_{\rm Pl}$ is the gravitino mass, and we introduced the parameter $\omega_R$ which fulfils the inequality in Eq.~\eqref{eq:irred}
\cite{Dine:2009sw,Bellazzini:2016xrt}.
Eq. \eqref{eq:irred} imposes a lower bound on the R-axion mass which we will take into account when discussing the SUSY embedding of our scenario.

The R-axion is naturally heavy and unstable, and it can couple with RHN in SUSY extensions of the SM (we will show an explicit realisation in Appendix \ref{app:SUSY_model}).
In Section \ref{sec:gravitino}
we will discuss the phenomenological implications of the SUSY scenario taking into account both the gravitino abundance and the matter-antimatter asymmetry induced by the R-axion decay.

\subsection{ALP production}

Thermal scatterings with gluons and top quarks in the primordial plasma unavoidably produce a population of hot ALP.  Throughout this paper we assume that non-thermal production mechanisms, like misalignment, yield subleading contribution to the ALP abundance. As a rough requirement for this, we set the ALP decay constant greater than the reheating temperature of the Universe: $f_{a} > T_{\rm RH}$.

The evolution of the abundance of axions can be traced using Boltzmann equations,
and it has been deeply investigated in \cite{Salvio_2014,Bouzoud:2024bom}.
For the  inflaton field $\Phi$ with energy density $\rho_\Phi$ and decaying with rate $\Gamma_\Phi$,  they take the compact form 
\begin{align}
\label{eq:Boltzmannequationcosmological}
        HZ z' \frac{d \rho_{\Phi}}{dz'} &= -3H \rho_{\Phi}- \Gamma_{\Phi} \rho_{\Phi} \, ,
        \qquad 
        sHZz' \frac{d Y_{a}}{dz'} = 3sH(Z-1) Y_{a} + \gamma_{\rm prod}  \bigg(1- \frac{Y_{a}}{Y^{ \rm eq}_{a}}\bigg) ,
        \\
        \nonumber 
        z'&=T_{\rm RH}/T, \quad (Z-1) \equiv - \frac{\Gamma_{\Phi}\rho_{\Phi}}{4H \rho_{R}} , \quad  Y_{a} \equiv n_{a}/s, \quad Y_{a}^{\rm eq}  \simeq 2.15 \times 10^{-3}, \quad  n_{a}^{\rm eq}= T^{3}/ \pi^2 \, ,
\end{align}
where $H$ is the Hubble expansion rate and $Y_{a}^{\rm eq}$ is the equilibrium Yield of the ALPs considering the Maxwell-Boltzmann statistics.
 $\gamma_{X_i \to a}$ are the thermal ALP production rates, where $X_i$ particles are in the thermal bath, namely SM particles and RHNs. 
 The  dominant production channels are the top and the gluon ones,
 and the associated production rates can be 
 written as 
 \cite{Salvio_2014}
\begin{equation}
\label{eq:ALPproductionrate}
    \gamma_{\rm prod} \equiv \sum_i\gamma_{X_i \to a} \approx \kappa \frac{T^{6}}{f_a^2}, \qquad \kappa \equiv \frac{ \zeta (3)}{(2 \pi)^5 } \bigg[ 37 C_{t}'^{2} y_{t}^2 + 8 C_{g}^{2} \alpha_{s}^2 F_{3}\bigg(\frac{m_{g}}{T} \bigg) + \text{subleading sources} \bigg] \, ,
\end{equation}
where $F_3(m_g / T)$ is a function that parameterizes the ALP production rate due to gauge interactions that can be found in~\cite{Salvio_2014}\footnote{The latest evaluation of ALP production via its gluon coupling~\cite{Bouzoud:2024bom} finds a rate smaller than~\cite{Salvio_2014} by a factor of a few, but as our ALP is dominantly produced via its top coupling this does not have a significant impact on our predictions.
}.
$C_{t}', C_{g}$ encode the dependence on the UV completion of the axion couplings. 
It is important to notice that $C_t'$ appearing in \eqref{eq:ALPproductionrate} is a combination of the couplings in \eqref{eq:axionl}, obtained after an ALP dependent chiral rotation of the SM fermions (see \cite{Salvio_2014} for details), concretely $C_t' = C_{Q_3}-C_{t}$. It controls the 
resulting ALP phase 
in front of the top Yukawa coupling.
For our purposes all the SM Yukawa couplings are negligible compared to the top one, hence we can neglect them in ALP production.
We conventionally set $C_{t}' = C_{g} = 1$ and numerically we have $\kappa \approx 5 \times 10^{-3}$.
The $X_i \to a$ processes that produce ALPs from initial states containing $X_i$ are dominantly $2\to2$, for example $gq \to qa$ or $Q_3 \phi^* \to a \bar{t}_R$.

In addition to the standard ALP-SM couplings studied in~\cite{Salvio_2014}, in our scenario we have the coupling between the ALP and the RHN,
and we should inspect if this will alter the production mechanisms of the ALP in the early Universe.
Let us first comment on the production from the thermal RHN sector via $N \, N \to a$ 
with width $\Gamma_{NN \to a} \propto \big(M_N/f_a\big)^2 m_a/8 \pi$. Comparing this with the ALP production via top Yukawa coupling, we find that parametrically
\begin{equation}
\frac{\Gamma_{\phi Q_3 \to t a}}{\Gamma_{N N \to a}} \propto y_{t}^2 \frac{T^3}{M_{N}^2 m_a} \bigg|_{T = T_{\rm RH}} \gg 1 \, ,
\end{equation}
 since $T_{\rm RH} \gg M_N \sim m_a$ in all the parameter space we consider.
In addition, the ALPs can be also produced via scatterings involving its coupling with the RHN, e.g. $N \, N \to a \ a $ via t-channel, or s-channel $N \, N \to a \ a $ with ALP trilinear vertex, but these are suppressed with an extra power of $1/f_a$. 
 Finally notice that the derivative ALP coupling to RHNs of Eq.~(\ref{eq:axionl}), upon performing an ALP-dependent rotation of the SM fields, induces not only the mass coupling $a\overline{N_R}N_R$, but also the Yukawa one $a L \phi N_R$, where $\phi$ is the Higgs field. This contributes to scattering processes
with the bath like
$\phi \, N \to L a $, but these
are suppressed by the smallness of the RHN Yukawa coupling. We conclude that in our scenario the ALP abundance is still determined by the ALP couplings with gluons and tops.

\begin{figure}[h!]
\centering
\includegraphics[width=12cm]{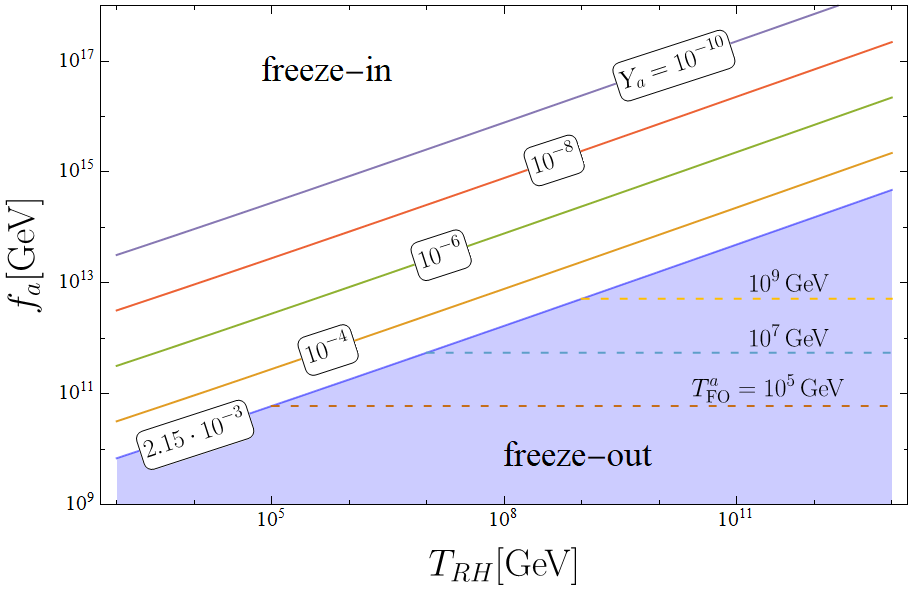}
\caption{\small{Parameter space $f_a - T_{\rm RH}$ of the cosmological ALP yield. In the blue area, the ALP yield is fixed by freeze-out computation with $Y_{a} = Y_{a}^{\rm eq} \simeq 2.15 \times 10^{-3}$. Dashed lines show different values of decoupling (i.e. freeze-out) temperature $T_{\rm FO}^a$. Conversely, in the white region, the production is controlled by the freeze-in mechanism with $Y_a = r Y_a^{\rm eq}$, represented by the contour lines.}}
 \label{fig:freezeoutin}
\end{figure}

 The solution to the BEs~\eqref{eq:Boltzmannequationcosmological} for the ALP abundance at $T \ll T_{RH}$ is\cite{Salvio_2014}
\begin{equation}
    \frac{Y_{a}}{Y_{a}^{\rm eq}} = \Big(1+r^{-3/2}\Big)^{-2/3} \simeq
    \begin{cases}
       \,  r \hspace{2mm}\text{for}\hspace{2mm} r \ll 1\\
        \, 1 \hspace{2mm}\text{for}\hspace{2mm} r \gg 1
    \end{cases} \, ,
\end{equation}
with 
\begin{equation}
    r \equiv \frac{2.4}{Y_{a}^{\rm eq}} \frac{\gamma_{\rm prod} }{H s} \bigg|_{T = T_{\rm RH}} \approx   1.4 \times 10^4 \times \kappa\frac{T_{\rm RH}}{10^{7}\text{GeV}} \bigg(\frac{10^{11} \text {GeV}}{f_{a}}\bigg)^{2}   \, .
\end{equation}
Thus the ALP, once it is produced, can either be in thermal equilibrium with the plasma ($r \gg 1$), with abundance given by $Y_{a} = Y_{a}^{\rm eq} \simeq 2.15 \times 10^{-3}$, or have a lower abundance  $Y_{a} \approx r Y_{a}^{\rm eq}$ ($r \ll 1$), i.e. the ALP is frozen in (FI). In the case of freeze-out (FO), the ALPs thermalise with the SM and have the same temperature as the bath. In Fig.~\ref{fig:freezeoutin}, the parameter space of $f_a-T_{\rm RH}$ for ALP's freeze-out and freeze-in is presented, where the white area corresponds to the FI and the blue region to the FO, with constant yield $Y_a = Y_a^{\rm eq}$. 
In the FI area (in white) the yield value is displayed with colored lines, and it gets reduced when increasing $f_a$ simply because the ALP-SM couplings are reduced.
In the FO region (in blue) the ALP yield is fixed, and
we show
the freeze-out temperature $T_{\rm FO}^a$ between the bath and ALPs
(dashed lines), which only depends on the value of $f_a$. 

In the numerical computations in the following, we will typically 
consider two benchmark scenarios: 1) the FO case with $Y_{a}^{\rm eq} \simeq 2.15 \times 10^{-3}$, and 2) the FI one with $Y_{a} = 10^{-4}$.

\subsection{ALP decay}

Depending on the regions of the parameter space, 
some abundance of ALP is produced by thermal scatterings (FO or FI). 
After those interactions are no longer effective in the early universe, the ALP will start decaying, mostly to SM, to the RHN and possibly to SUSY particles. The different decay rates of the ALP into SM and RHN particles are $a \to gg$ to gluons, $a \to NN$ to the RHNs, $a \to t\bar t$ to the tops or possibly, if the model contains gravitini, $a \to \tilde G \tilde G$, to the gravitini themselves. While the latter decay mode is always subdominant with respects to decays into RHNs, it is important to determine the resulting gravitino abundance, as we will discuss in Sec.~\ref{sec:gravitino}.
The rates of ALP decay channels are presented in Table \ref{tab:channels}~
\cite{Bellazzini:2017neg, Hamada_2014}. 

\begin{table}[]
\begin{center}
\begin{tabular}{lll}
\hline
\multicolumn{1}{|l|}{Channel} & \multicolumn{1}{l|}{$\Gamma_{\rm channel}$} & \multicolumn{1}{l|}{Minimal realisation}  \\ 
\hline
\multicolumn{1}{|l|}{$a \to gg$} & \multicolumn{1}{l|}{$C_{g}^2 \frac{\alpha_{s}^2}{32 \pi^3}\frac{m_{a}^3}{f_{a}^2}$} & \multicolumn{1}{l|}{Yes}  \\ 
\hline
\multicolumn{1}{|l|}{$a \to NN$} & \multicolumn{1}{l|}{$\frac{m_{a} M_{N}^2}{{8 \pi f_{a}^2}} \sqrt{1-4 \frac{M_{N}^2}{m_{a}^2}}$} & \multicolumn{1}{l|}{Yes} 
\\ 
\hline
\multicolumn{1}{|l|}{$a \to t \bar t$} & \multicolumn{1}{l|}{$C_{t}^2 \frac{m_{a} m_{t}^2}{{8 \pi f_{a}^2}} \sqrt{1-4 \frac{m_{t}^2}{m_{a}^2}}$} & \multicolumn{1}{l|}{Yes} 
\\ 
\hline
\multicolumn{1}{|l|}{$a \to \tilde G \tilde G$} & \multicolumn{1}{l|}{$\frac{1}{4\pi} \frac{m_a^5 \omega_R^2}{f_a^2 F^4}$} & \multicolumn{1}{l|}{No (only in SUSY models)} 
\\  
\hline
\end{tabular}
\caption{\small{Here we tabulate the expression of the decay rates of the ALP $a$. In the last column, we mention if the channel belongs to the minimal realisation of the model as exemplified in Eq.\eqref{eq:axionl}. } }
\label{tab:channels}
\end{center}
\end{table}

As we already anticipated in the introduction, our mechanism enhances the leptogenesis yield by inducing a non-thermal population of RHNs coming from $a \to N \, N$. In this sense, other decay channels like $a \to t \bar{t}, gg$ are a \emph{loss} for ALP leptogenesis. Looking at table \ref{tab:channels}, it appears that this constrains the ratio $m_a/M_N$ and the value of $M_N$. Requiring as an example that $\text{Br}_{a \to NN} > 1/2$, induces roughly $m_a/M_N \lesssim 2 \pi/\alpha_s \approx 4.5/\alpha_s$, for $C_g =1$ and $M_N \gg m_t$.

\medskip

Throughout this paper, we will use the benchmark value $m_a/M_N =3$, for which the dominant decay channel in the parameter space of interest turns out to be $a$ $\rightarrow$ $NN$. This sets the temperature at ALP decay which is thus well approximated by solving the relation 
\bea 
H(T =T^a_D) = \Gamma(a \rightarrow N \, N ) \, , 
\eea 
which implies 
      \begin{equation}
            T_{D}^{a} = \frac{M_{N}}{f_{a}}\sqrt{ \frac{m_{a}}{8 \pi} \frac{M_{\rm Pl}}{1.66 \sqrt{g_{*}}} \sqrt{1-4 \frac{M_{N}^2}{m_{a}^2}} } \,,
    \end{equation}
with $g_*=106.75$.

It might be surprising that the axions are mainly produced via their coupling with tops, while eventually they decay mostly to RHNs. This can be understood by the fact that ALP production channels are scatterings, and those involving RHNs (including the one induced by the coupling $a L \phi N_R$ that appears after a chiral rotation) are suppressed by a much smaller coupling with respect to those involving tops.
Also, the scattering rates scale like $T^6$ while the decays and inverse decays only scale like $m_a$ or $m_a^3$. At $T=T_{\rm RH} > m_a$, the scattering dominates due to large temperature while for temperatures close to $T^a_D < m_a$, the scatterings are negligible. 

With those tools in hand, we can now address the mechanism of ALP-induced leptogenesis.

\section{Leptogenesis via ALP decay}
\label{sec:leptogenesis}

In the previous section, we studied the production and the decay of the ALP. Now, after a concise review of standard and resonant leptogenesis, we study the non-thermal leptogenesis mechanism coming from the new population of RHNs, induced by the ALP decay at $T \sim T_D^a$.

\subsection{Reminder on standard and resonant leptogenesis}
Leptogenesis 
was first
proposed in \cite{FUKUGITA198645} as an elegant mechanism capable of achieving successful baryogenesis via the out-of-equilibrium and $CP$-violating decay of a RHN, which is coupled to the lepton doublet $L$ and the Higgs boson $\phi$ through the complex Yukawa coupling $y_{\nu}$
\begin{equation}
\label{eq:see}
    \mathcal{L} =  - y_{\nu} \overline{L} \cdot \tilde{\phi} N - \frac{1}{2} M_{N} \overline{N}^c N +h.c.
\end{equation}
 where $M_N$ is the Majorana mass of the RHN. The Lagrangian of Eq.~(\ref{eq:see}) gives rise to neutrino masses via the standard type-I see-saw mechanism~\cite{Minkowski:1977sc,Yanagida:1979as,Mohapatra:1979ia,Schechter:1980gr}, and we impose the requirement to reproduce the observed ones throughout our analysis.
 
 The RHN decays are able to generate dynamically a lepton asymmetry, which is partially converted into a baryon asymmetry through Sphaleron processes, active in the temperature range $T \in [10^2, 10^{12}]$ GeV. The  baryon asymmetry originating from leptogenesis can be parameterized in the following way 
 \begin{equation}
\label{eq:YB}
    Y_{\Delta B}
    \simeq Y_N^{\rm eq}\,c_{\rm sph}\,\epsilon_{\rm CP}\, \kappa_{\rm wash}\,, \qquad \epsilon_{\rm CP} \equiv \frac{\Gamma(N\to L \phi) - \Gamma(N\to  \overline{L}\, \overline{\phi})}{\Gamma(N\to L \phi) + \Gamma(N\to  \overline{L}\, \overline{\phi})}  \, ,\qquad c_{\rm sph} = \frac{28}{79}
\end{equation}
where $Y_N^{\rm eq} \equiv  n_N^{\rm eq}/s \approx 2.05g_N/g_{\star,s}$ is the equilibrium yield of RHN at $ T\gg M_N$, $c_{\rm sph}$ is the sphaleron conversion factor, $\epsilon_{\rm CP}$ is the $CP$-asymmetry parameter and $\kappa_{\rm wash}$ accounts for washout effects, i.e. active processes that erase the created lepton asymmetry and accounts for the fact that only a fraction of the initial abundance of RHNs indeed decays out-of-equilibrium.
Even if $\kappa_{\rm wash}$ cannot be obtained analytically in general, we can distinguish two qualitative regimes: the strong wash-out regime with  $\kappa_{\rm wash} \ll 1$ and the weak wash-out regime $\kappa_{\rm wash} \sim 1$. The rough regime of a specific model can be obtained by evaluating the parameter
\begin{equation}
K \equiv \frac{\Gamma_D}{H(T=M_N)} \simeq \frac{\tilde m_{\nu}}{10^{-3} \text{eV}} \, , 
 \qquad \Gamma_D = \Gamma(N\to L \phi) + \Gamma(N\to  \overline{L}\, \overline{\phi}) \, ,
\end{equation}
where $\Gamma_D$ is the total decay width and $\tilde m_{\nu}$ is the effective light-neutrino mass \cite{Plümacher1997549}. A value of $K$ much larger than 1 means that the inverse decay are still active when the population of RHNs starts to be exponentially suppressed, leading to a strong wash-out and small $\kappa_{\rm wash} \ll 1$. Conversely,  $K$ close to 1 means that the inverse decays are nearly decoupled and  $\kappa_{\rm wash} \sim 1$. Therefore, considering the experimental measurements of the neutrino mass scale ($ 8 \cdot 10^{-3}$ eV $\simeq m_{\rm sol} \lesssim \tilde m_{\nu} \lesssim m_{\rm atm} \simeq 0.05$ eV \cite{Gonzalez_Garcia_2021}), the washout is typically $K \simeq (10 \div 50)$, which implies $\kappa_{\rm wash} \simeq (10^{-2} \div 10^{-3})$ (see ref.\cite{Buchm_ller_2005} for a detailed computation). This corresponds to the strong washout regime, indeed $\kappa_{\rm wash} \ll 1$\footnote{However, a weak washout scenario with $K \lesssim 1$, i.e. $\kappa_{\rm wash} \simeq 1$, is achievable within thermal leptogenesis at the price of a tuning in the parametrization matrix\cite{Moffat:2018wke, Granelli:2021fyc}. }.

Those considerations allow us to derive consequences on the scale of leptogenesis. For the illustration, let us consider the standard thermal leptogenesis scenario with hierarchical RHN masses, i.e. $M_1 \ll M_{2,3}$. In this case, only the lightest neutrino $N_1$ contributes to generate the final BAU. The $CP$-asymmetry parameter is bounded
from above \cite{Davidson_2002} for a hierarchical spectrum of RHN: $|\epsilon_{\rm CP}| \lesssim (3M_N (m_{\nu_3} - m_{\nu_1} 
))/(8\pi v_{\rm EW}^2
) $
where $v_{\rm EW} = 246$ GeV is the SM Higgs vacuum expectation value (vev). Plugging this upper bound in Eq.\eqref{eq:YB} and requiring to match Eq.\eqref{eq:observed_BAU} we obtain the Davidson-Ibarra bound $M_{N_1} \gtrsim 10^{11} (10^{9})$ GeV in strong (weak) washout regime \cite{Davidson_2002, Hamaguchi_2002}.

The Davidson-Ibarra bound implies that even in the most optimistic case of weak wash-out, the scale of 
 the see-saw lies much beyond the reach of detection of colliders.  One needs however to notice that the masses $M_{1,2,3}$ enters mostly via the $CP$-violating parameter $\epsilon_{\rm CP}$, which contains contributions from the self-energy correction and the vertex correction. The self-energy correction  contains a resonance due to the $N_i$ propagator $1/|M_1^2 - M_i^2|$, with $i \neq 1$, which allows to greatly enhance $\epsilon_{\rm CP}$, at the price of tuning the masses of the heavy neutrinos to be quasi-degenerate. 
 In so doing, one can lower the value of the RHN masses down to TeV-scale by considering resonant leptogenesis (see \cite{Pilaftsis:2003gt, Pilaftsis_2005, Granelli:2020ysj} for reviews). The generalised $CP$-violating parameter takes the form\cite{Pilaftsis_2004,Davidson:2008bu}:
\begin{equation}
\label{eq:res}
    \epsilon_{1} = \sum_{i = 1,2}\frac{\text{Im} [(y_{\nu}^{\dagger} y_{\nu})^2_{1i}]}{(y_{\nu}^{\dagger} y_{\nu})_{11} (y_{\nu}^{\dagger} y_{\nu})_{ii}} \frac{(M_1^2 -M_i^2) M_1 \Gamma_{N_i}}{(M_1^2 -M_i^2)^2 + M_1^2 \Gamma_{N_i}^2} \, ,
\qquad 
    \Gamma_{N_i}= \frac{(y_{\nu}^{\dagger} y_{\nu})_{7ii}}{8 \pi} M_{i}
\end{equation}
where $\Gamma_{N_i}$  is the decay width of $N_i$ at tree-level. In the regime of maximal resonance $(M_1^2 -M_i^2)^2 \sim M_1^2 \Gamma_{N_i}^2 $, the $CP$-violating parameter is enhanced parametrically to 
\bea 
\epsilon_{1} \sim \sum_{i = 1,2}\frac{\text{Im} [(y_{\nu}^{\dagger} y_{\nu})^2_{1i}]}{(y_{\nu}^{\dagger} y_{\nu})_{11} (y_{\nu}^{\dagger} y_{\nu})_{ii}}  \, ,
\eea 
which is not suppressed by small couplings. This resonant effect is not only present in the decay of heavy neutrinos but also in the scatterings, as we discuss in Appendix \ref{app:CP_viol_scat}
following the lines of \cite{Pilaftsis:2003gt, Nardi:2007jp}. 
Resonant leptogenesis is  a very fruitful framework to enhance the yield of leptogenesis and make it more detectable. 

In the scenario we described so far, the heavy neutrino follows a thermal population, which is produced via inverse decays and  thermal scatterings with the plasma. We can now move to our proposal.

\subsection{ALP leptogenesis}
\label{sec:alp_lepto}
Having reviewed standard thermal leptogenesis, we now present our new proposed mechanism of `ALP leptogenesis'. As anticipated in the introduction, the idea is to source an out-of-equilibrium population of RHNs from ALP decays. As we will see, this will soften washout effects and enlarge the parameter space of successful leptogenesis.
In order for our mechanism to work in generating eventually a baryon asymmetry,
there are two requirements to impose. First of all, we remind that the decay must be kinematically allowed: $m_a > 2 M_N$. Then, as the asymmetry is produced in the lepton sector, it needs to be transferred to the baryon sector via the sphalerons, which  efficiently convert the  asymmetry if the decay of neutrinos has happened mostly before the electroweak phase transition:
    \begin{equation}
    \label{eq:sphaleron_condition}
    T_{D}^{a} \gtrsim T_{\rm sph}^{\rm dec} \simeq 132 \,  \text{GeV}  \qquad \text{(Sphalerons are active at the moment of the ALP decay)}\, .
    \end{equation}
 This is an effective minimal setting for successful baryogenesis. A possible advantage of our mechanism is to suppress the wash-out and to obtain $\kappa_{\rm wash} \approx 1$. This occurs if the inverse decay  of RHNs is decoupled at the time of the decay of the ALP to the RHN, which is approximately given by:
    \begin{equation}
    \label{eq:firstconstr}
         T_{D}^a \lesssim \frac{M_{N}}{20} \qquad \text{(RHN inverse decays are decoupled for ALP leptogenesis)} \, .
    \end{equation}
Hence, $T^a_D$ has to lie in the range 
\bea 
\label{eq:window}
132 \,  \text{GeV} \lesssim T_{D}^a \lesssim \frac{M_{N}}{20} \, . 
\eea 
The inequalities in Eq.~\eqref{eq:window} select a window in the parameter space where the mechanism of ALP leptogenesis could work and be more efficient than thermal leptogenesis. 
The resulting interesting parameter space in the ALP mass vs decay constant is displayed in Figure \ref{fig:dil1}.
In the left panel, we fix the reheating temperature to be large enough for the ALP abundance to be set by the freeze-out mechanism. 
The viable region is delimited by an orange line (for the sphaleron condition \eqref{eq:sphaleron_condition}) and by a blue line (for the condition \eqref{eq:firstconstr}).
In the right panel of Figure \ref{fig:dil1}, we instead set the reheating temperature such that the
freeze-in abundance of the ALP is $Y_a  = 10^{-4}$.
The viable region for ALP leptogenesis,
corresponding to the window \eqref{eq:window}, is again delimited by orange and blue lines.

We further note that
the condition in \eqref{eq:firstconstr} combined with $M_N < m_a /2$
automatically implies that the ALP
shall decay at the temperature $T_D^a < m_a/40$ much smaller than $m_a$. At decay, the ALPs are thus essentially ``at rest'' in the plasma frame. The decay products (mostly RHN in our case) then share typical energies $E_N \sim m_a/2$ and are 
far away from equilibrium. From this consideration, the boost of RHNs with respect to the plasma frame is simply $\gamma_E \sim m_a/(2 M_N)$.
Leptogenesis will now proceed via the out-of of equilibrium decay of RHNs. 

Our scenario involves however several subtleties with respect to the usual thermal leptogenesis, subtleties that we will now enumerate.

\subsection{Kinetic non-equilibrium and early matter domination}
\label{sec:subtle}
In this subsection, we go over the main non-standard features of our leptogenesis mechanism: first, the RHNs which are emitted from the ALP decay never reach kinetic equilibrium and secondly, the late decay of the ALP can lead to an early matter domination phase, which would dilute the baryon abundance when the ALP decays.

\paragraph{Kinetic equilibrium is never reached:} 

In the scenario we are considering, the ALP decouples from the plasma and, afterwards, it decays into the RHNs.
For what concerns the RHN population, their momentum distribution is inherited from the ALP decay and it is different from the equilibrium kinetic distribution: since $T_{D}^{a} <  m_a/40$, ALPs decay almost at rest and the kinetic distribution of RHNs is expected to be peaked around $E_N \sim m_a/2$. Thus the RHNs are produced with typical boost factor $\gamma_E \sim m_a/(2 M_N)$ and are thus not in kinetic equilibrium. 

After the emission of the RHN,
they could attain kinetic equilibrium with the bath via particle-number conserving $2 \to 2$ interactions with species in the thermal bath like $N \phi \to N \phi $, $Nt \to Nt$. However the rate of such scatterings scale like $y_{\nu}^4$, 
\bea 
\Gamma_{2 \to 2} \propto \frac{y_{\nu}^4 T}{8\pi}, \qquad y_{\nu}^2 \approx \frac{m_\nu M_N}{v_{EW}^2} \sim 2\times  10^{-12} \frac{M_N}{\text{TeV}} \, , 
\eea  
and, for $M_N/20 \sim T_{D}^{a} \sim 1-10 \text{ TeV}$, are strongly decoupled.
In particular, $\Gamma_{2 \to 2}/H \approx 10^{-16} \cdot (M_N/(10^8~\text{GeV})) \cdot M_{\rm Pl}/T$, so that number conserving $2 \to 2$ interactions are never in equilibrium for $M_N \lesssim 10^8~\text{GeV}$, where we use and check that $T >$ (or $\gg$) $T_{\rm sph}^\text{\rm dec}$. While larger values of $M_N$ are not of interest for our work, we comment that values increasingly larger than $10^8~\text{GeV}$ demand an additional upper limit on $f_a$, to realise RHNs decay at $T >$ (or $\gg$) $T_{\rm sph}^\text{\rm dec}$, which reaches $f_a \lesssim 10^{16}$~GeV at $M_N \simeq 4\times 10^{11}$~GeV. To summarise, in the entire parameter space of interest for ALP leptogenesis, RHN decay away from kinetic equilibrium.

    \paragraph{The axion domination dilutes the baryon yield:} In addition, $T_{D}^a \ll m_a$ because we are working with $T_{D}^a \lesssim M_N/20$ to avoid $N$ inverse decays (see Eq.~(\ref{eq:firstconstr})), and $M_N < m_a/2$ to possibly have ALPs decays to RHNs. Since furthermore the decay of the ALP is very slow
 with respect to the timescale of the leptogenesis, the ALPs, after becoming non-relativistic, may temporarily dominate the energy density of the Universe. This eventually results in a dilution of the leptogenesis yield due to the injection of entropy coming from the ALP decay.
The fact that the decay of the dominating ALP dilutes the abundance of \emph{its own} decay products may seem counter-intuitive at first sight. To convince oneself that this is indeed the case, one may notice that the decaying ALPs do constitute an early-universe relic, and therefore they should also be diluted by entropy injections as they decay, like all other relics.
We confirm this result numerically in the Appendix \ref{app:dilution}, and note that it is also confirmed in another context by the results of~\cite{Nemevsek:2022anh}.

We can define the dilution factor as the ratio of comoving entropy $S= s a^3$ after and before the decays
\begin{equation}
\label{eq:DSMentropyratio}
    D_{\rm SM}  \equiv \frac{S_{\rm SM}^{\rm after}}{S_{\rm SM}^{\rm before}} \, .
\end{equation}
 Following~\cite{Cirelli_2019}, the dilution factor can be estimated as
\begin{equation}
\label{eq:DSMest}
    D_{\rm SM}  = \bigg[1 + \bigg(0.43\times \frac{Y_a}{Y_a^{\rm FO}}g_{\star}^{1/4}\frac{g_{a}}{g_{*}} \frac{m_{a}}{\sqrt{\Gamma_{a} M_{\rm Pl}}} \bigg)^{4/3} \bigg]^{3/4} \,,
\end{equation}
where $g_{a} = 1$, $Y_a$ is the ALP yield and $Y_a^{\rm FO}  \simeq 2.15 \times 10^{-3}$ is the freeze-out abundance of the ALP. In Appendix \ref{app:dilution}, we verify numerically that this analytical estimate is accurate. We comment that for dilution factor smaller than $D_{\rm SM} \lesssim 50$, the disagreement remains below 15 \%. 
We then study the features of the matter-dominated era induced by the ALP. 

\begin{figure}[h!]
\centering
\centering
{\includegraphics[width=0.49\textwidth]{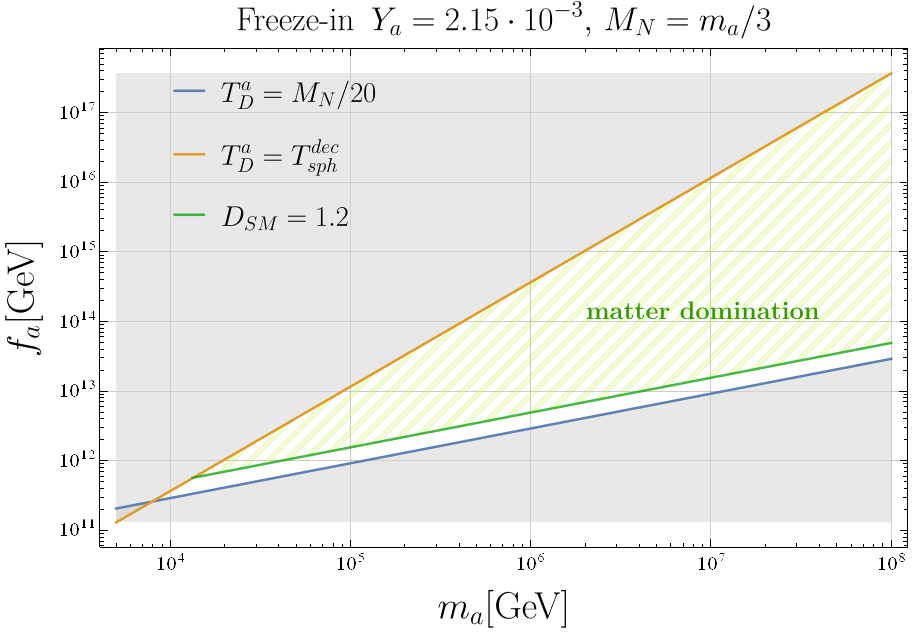}}
\hspace{1mm}
{\includegraphics[width=0.49\textwidth]{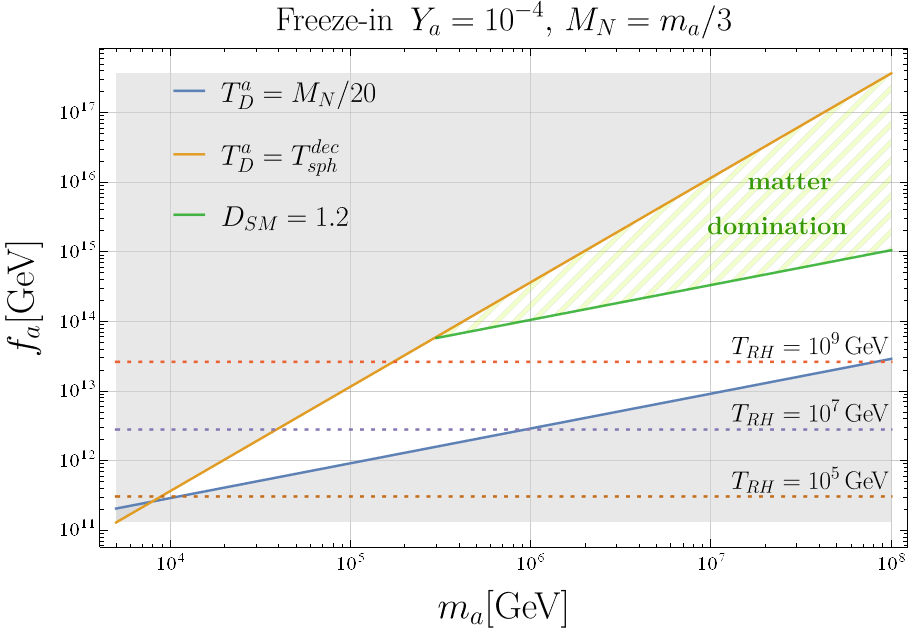}}\\
\caption{\small{
ALP decays in the early Universe lead to successful leptogenesis outside the gray area, delimited by the requirements of active sphalerons (orange line) and limited wash-out of the asymmetry (blue line). 
The region of successful ALP leptogenesis contains a green hatched area, where the ALP additionally induces an early phase of matter domination: while the BAU it still successfully reproduced there, the gain with respect to non-ALP leptogenesis is less significant than in the white area.
In the left panel, the ALP is mainly produced via the top Yukawa coupling and it is frozen out with $Y_{a} = 2.15 \times 10^{-3}$. In the right panel, the dashed lines show the reheating temperature values required to freeze-in the ALP with $Y_{a} = 10^{-4}$. We set $M_N = m_a/3$ as benchmark. See text for more details.}}
 \label{fig:dil1}
\end{figure}

 We can estimate the region of the parameter space where the dilution induced by the matter-dominated period will be significant in the final abundance of RHNs (and hence of the $B-L$ asymmetry).
 In Fig.~\ref{fig:dil1}, we consider $D_{\rm SM} = 1.2$ as an indicative threshold for the ``large'' dilution region and represent this region in green. After the decay of the ALP, the RHN yield is significantly diluted by $Y_N^{\rm after}= \frac{ Y_N^{\rm before}}{D_{\rm SM}}$. In the white region, the ALP does not dominate the energy content of the Universe and the resulting abundance for the Majorana neutrinos produced via ALP decays is $Y_{N} = 2 Y_{a}$ at $T = T_{D}^{a}$.


So far we isolated the region of parameter space where the enhancement of leptogenesis is likely to occur. We now turn to the numerical study of leptogenesis.

\subsection{Basics of our numerical analysis}

In this subsection, we present the main numerical expressions we solved to obtain our results. 
We can describe the dynamics of the process via Boltzmann equations (BEs). Eventually, the results will be discussed in Sec.~\ref{sec:results_discussion} and an analytical formula for the final baryon yield will be presented. 

\subsubsection{System of BEs}
In our scenario, there is a first thermal population of RHNs generated via 
interactions in the thermal bath and a second one, which is non-thermal and is produced via ALP decays. 
The non-thermal population produced through ALP decays 
is boosted by a factor 
$\gamma_E \sim m_a/(2M_N)$
and some care will be needed to derive the corresponding Boltzmann equation.


In practice, the two populations will coexist for a very limited amount of time since one of the requirements of our mechanism is that RHN inverse-decays are decoupled at the time of ALP decay (see \eqref{eq:firstconstr}).  
As a consequence, we can essentially solve two sets of different BEs for the two abundances in two different time intervals.

\paragraph{$z < z_{\rm matching}$: thermal contribution}
We first solve the usual set of coupled BEs for standard thermal leptogenesis
\begin{equation}
\label{eq:thermalBE}
\text{(BE for thermal leptogenesis)} \qquad 
    \begin{cases}
         \frac{d Y_{N}}{dz} = - \frac{\gamma_{D}}{H s z} \big(\frac{Y_{N}}{Y_{N} ^{\rm eq}}-1\big)\\
         
        \frac{d Y_{B-L}}{dz} =  \frac{\gamma_{D} \epsilon_{\rm CP}}{H s z} \big(\frac{Y_{N}}{Y_{N} ^{\rm eq}}-1\big) - \frac{\gamma_D}{Hsz} \frac{Y_{B-L}}{2 Y_{l}^{eq}}
    \end{cases}
\end{equation}
where $z=M_N/T$, $H=H(z)$ is the Hubble parameter at temperature $T$ in a radiation-dominated Universe, $Y_{N}^{\rm eq}$ and $Y_{l}^{\rm eq}$ the equilibrium Yields of RHN and leptons considering the Maxwell-Boltzmann statistics, $\gamma_D$ is the reaction density of the RHN decay $N \rightarrow L \,  \phi$.
 We use this set of BEs until the $z_{\rm matching} = 10$, when the thermal RHN decayed and this abundance essentially vanishes. 

\paragraph{$z > z_{\rm matching}$: ALP decay contribution}
The $B-L$ asymmetry is then matched on a second set of BEs that accounts for the decay of the ALP, which we present now. 
Before presenting the BEs, we comment on the kinetic equilibrium of the species involved in the process. As we have seen in section \ref{sec:subtle}, the RHN from ALP decay cannot reach kinetic equilibrium and their energy is very peaked around $E_N \approx m_a/2$. Consequently we cannot follow the usual computation of the integrated Boltzmann equations. 

 We can now work out the equations for the abundance of RHN, $n_N$, 
 starting from the un-itegrated form of the BE and following steps similar to the usual procedure
 of~\cite{Hahn-Woernle:2009jyb}.
 The relevant interactions for the RHN are the decay/inverse-decay into $L \phi$ and the ALP decay into $NN$.
 The former, even though not efficient enough to generate a sizeable RHN thermal population, dictates how the RHN abundance gets depleted.
 The latter, instead, acts as a source term which creates RHNs.
 The resulting BE reads:
\begin{equation}
\label{eq:nontherBE2}
    \frac{\partial n_{N}}{\partial z} =\frac{M_N \Gamma_D}{z H(z) } \bigg( \underbrace{\int \frac{d^3p_{N}}{E_{N}(2 \pi)^3} \overbrace{f^{\rm eq}_{N}}^{\approx 0}}_{\text{thermal abundance}} - \underbrace{\int \frac{d^3p_{N}}{E_{N}(2 \pi)^3}f^{}_{N} }_{
    \text{ALP decay abundance}
    }\bigg) +
    \underbrace{
     \frac{1}{ H(z) z} \int \frac{d^3 p_N} {(2 \pi)^3} C_{a\rightarrow N N}[f_{N}]
     }_{\text{ALP source}}
\end{equation}
where $\Gamma_D = y_{\nu}^2 M_N/8\pi$ is the RHN decay width.
The first term in Eq.\eqref{eq:nontherBE2} is the usual thermal abundance 
induced by inverse decay processes, 
and can be treated via well-known methods. 
In practice, for $z> 10$, this abundance is $\sim 0$. 
The last term is describing the source of the RHN abundance from ALP decay.
The second term instead governs how the RHN population is reduced through the decay into $L \phi$.
Since here most of the RHN population is originating from the source term (so from the ALP decay)
their distribution is unknown. On the other hand, we know that their distribution is sharply peaked around $E_N \simeq m_a/2$.
We can thus take the $E_N$ in the second term of Eq.\eqref{eq:nontherBE2} out of the integral and perform the integral.
We obtain
\begin{equation}
    \frac{\partial n_{N}}{\partial z} =\frac{ \Gamma_D}{z H(z) } \bigg( n^{\rm eq}_{N} - \frac{2 M_N}{m_a} n_{N} \bigg) + \text{ALP source}\quad \Rightarrow  \quad \frac{\partial Y_{N}}{\partial z} = - \frac{ \gamma_D}{ H(z) s z } \bigg( \frac{2 M_N}{m_a} \frac{Y_{N}}{Y^{\rm eq}_{N}} -1 \bigg)
    + \text{ALP source}
    \nonumber
\end{equation}
where the ratio of Bessel functions is reasonably approximated as $K_1(z)/K_2(z) \simeq 1$ given that $z>10$. We conclude that the fact that the RHNs produced via ALP's decays are not in kinetic equilibrium can be modelled in the BEs by the boost factor $2M_N/m_a$. 

Finally, the source term modelling the ALP decay $a \to N N$ in \eqref{eq:nontherBE2} can be rewritten taking into account the explicit expression for the collision term
\begin{equation}
\label{eq:nontherBE3}
      C_{a \rightarrow N N}[f_{N}]= \frac{1}{ E_{N}}  \int d\Pi_{N} d\Pi_{a} (2 \pi)^4 \delta^4(p_{N}+p_{N}-p_{a})  |\mathcal{M}_{a \rightarrow N N }|^2\frac{n_a}{n_a^{eq}} f_a^{eq}  \, 
\end{equation}
with phase space $\Pi_i = \frac{d^3 p_i}{2 E_i(2\pi)^3}$ for particle $i$ and squared matrix element $|\mathcal{M}_{a \rightarrow N N }|^2$.  In order to obtain Eq.\eqref{eq:nontherBE3}, we dropped the term for the inverse decay $N N \to a$ since its contribution is negligible at 
$z>z_{\rm{matching}}$,
and we considered the ALP to be in kinetic equilibrium, i.e. $f_a \simeq \frac{n_a}{n_a^{eq}} f_a^{eq}$. Moreover, we have included a factor of 2 in the expression for the collision term, since the ALP decays into two identical RHN 
(see e.g. \cite{D_Eramo_2021}).
We thus derive from \eqref{eq:nontherBE2} the following BE for the RHN

\begin{equation}
\frac{d Y_{N}}{dz} = - \frac{\gamma_{D}}{H s z} \bigg( \frac{2 M_N}{m_a}\frac{Y_{N}}{Y_N ^{\rm eq}}-1\bigg) + 2\frac{ \text{Br}_{N} \gamma_{a}}{H s z} \frac{Y_a}{Y_a ^{\rm eq}} \, ,
\end{equation}
where the reaction density $\gamma_a$ is
\begin{equation}
    \gamma_{a}  = s Y^{\rm eq}_{a} \frac{K_{1} \big(\frac{m_a}{T}\big)}{K_{2} \big(\frac{m_a}{T}\big)} \Gamma_{a} = \frac{T^{3} }{2 \pi^2} \bigg(\frac{m_a}{T}\bigg)^2 K_{1}\bigg(\frac{m_a}{T}\bigg) \Gamma_a = \frac{m_a^2 M_N K_1\big(z \frac{m_a}{M_N}\big) \Gamma_a}{2 \pi^2 z}
\end{equation}
with $\Gamma_a$ the total decay width of the ALP and
\begin{equation}
    Y_{a}^{\rm eq} = \frac{n_{a}^{\rm eq}}{s} = \frac{45}{4 \pi^4 g_{*}} \bigg(\frac{m_a}{T}\bigg)^2 K_2 \bigg(\frac{m_a}{T}\bigg) = \frac{45 m_a^2 z^2 K_2\big(z \frac{m_a}{T}\big)}{4 \pi^4 M_N^2 g_*} \, .
\end{equation}
Finally, putting all the pieces together, the BEs that track the evolution of the yields of the ALP, RHN and $B-L$ asymmetry in ALP leptogenesis (for $z>z_{\text{matching}}$) read:
\begin{equation}
\text{(BE for ALP leptogenesis)} \qquad 
\label{eq:ALPBEsystem}
    \begin{cases}
    \frac{d Y_a}{dz} = - \text{Br}_{N}\frac{\gamma_{a}}{H s z} \frac{Y_a}{Y_a ^{\rm eq}} - \text{Br}_{g}\frac{\gamma_{a}}{H s z} \big(\frac{Y_a}{Y_a ^{\rm eq}}-1\big) -\text{Br}_{t}\frac{\gamma_{a}}{H s z} \big(\frac{Y_a}{Y_a ^{\rm eq}}-1\big)\\
         \frac{d Y_{N}}{dz} = - \frac{\gamma_{D}}{H s z} \big( \frac{2 M_N}{m_a}\frac{Y_{N}}{Y_N ^{\rm eq}}-1\big) + 2\frac{ \text{Br}_{N} \gamma_{a}}{H s z} \frac{Y_a}{Y_a ^{\rm eq}}\\
        \frac{d Y_{B-L}}{dz} =  \frac{\gamma_{D} \epsilon_{\rm CP}}{H s z} \big(\frac{2 M_N}{m_a}\frac{Y_N}{Y_N ^{\rm eq}}-1\big) - \frac{\gamma_D}{Hsz} \frac{Y_{B-L}}{2 Y_{l}^{\rm eq}}
    \end{cases}
\end{equation}
where $Br_i$ are the branching ratios of the decay processes.
In this set of BEs~\eqref{eq:ALPBEsystem}, the equation for the ALP yield $Y_a$ has been added on top of the
BE for the RHN derived above.
It contains terms due to the interaction of ALPs with gluons, top quarks and RHNs. The first term models the ALP decay into RHNs: it has been derived similarly to the ALP source term in the BE for $Y_N$, neglecting the inverse decay $ N N \to a$, as we discussed in detail above. On the other hand, the other two terms, tracking the processes involving tops and gluons, can be easily computed following the standard derivation of the integrated BEs, since both tops and gluons are in thermal equilibrium. Actually, we checked that the ALP decays are the only relevant processes for $z>z_{\text{matching}}$, while the ALP inverse decays are ineffective at this stage.

\subsubsection{Example of solutions of the BE system}

We are now in position of solving the system of BEs. 
We solve the BEs for the thermal leptogenesis in Eq.~\eqref{eq:thermalBE} in the range $z \in [z_i,z_{\rm matching}]$ where $z_i = 10^{-3} \ll 1$, $z_{\rm matching} = 10$ and with initial conditions $Y_{N}(z_i) = Y_{B-L}(z_i)= 0$. This gives us values of $Y_{N}^{\text{therm}}(z=10)$ and $Y_{B-L}(z=10) \equiv Y_{B-L}^{\text{therm}}(z=10)$. The value of $z_{\rm matching}=10$ was chosen because this corresponds approximately to the end of the regime of thermal leptogenesis. We verified that modifying the matching value does not affect the final asymptotic $B-L$ asymmetry Yield, for any $z_{\text  {matching}} \in [10^{-3}, 20]$.
We then match the abundance and solve the BEs for the ALP leptogenesis in Eq.~\eqref{eq:ALPBEsystem} in the range $z \in [z_{\rm matching},100]$ with initial conditions $Y_a(z=10) = 2.15 \times 10^{-3}$, $Y_{N}(z=10) = Y_{N}^{\text{therm}}(z=10)$, $Y_{B-L}(z=10) = Y_{B-L}^{\text{therm}}(z=10)$.\\

\begin{figure}[h!]
\centering
\subfigure
{\includegraphics[width=9.2cm]{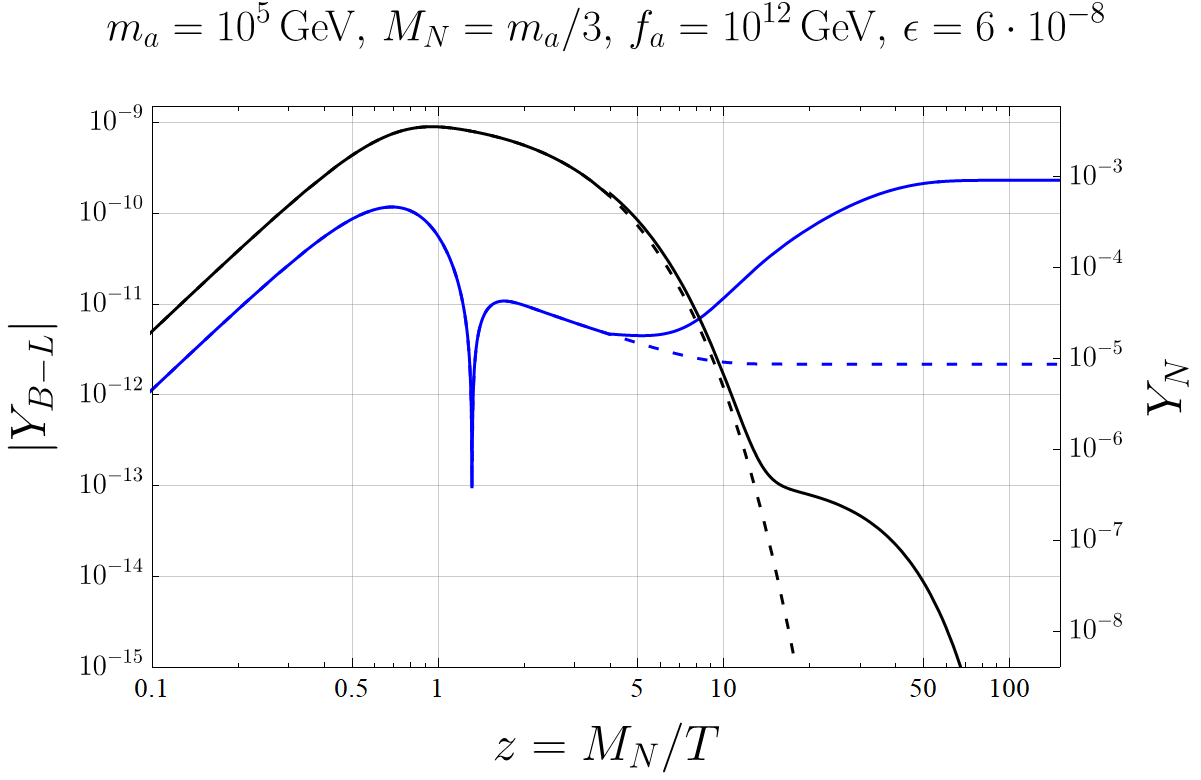}} 
\hspace{3mm}
\subfigure
{\includegraphics[width=8.2cm]{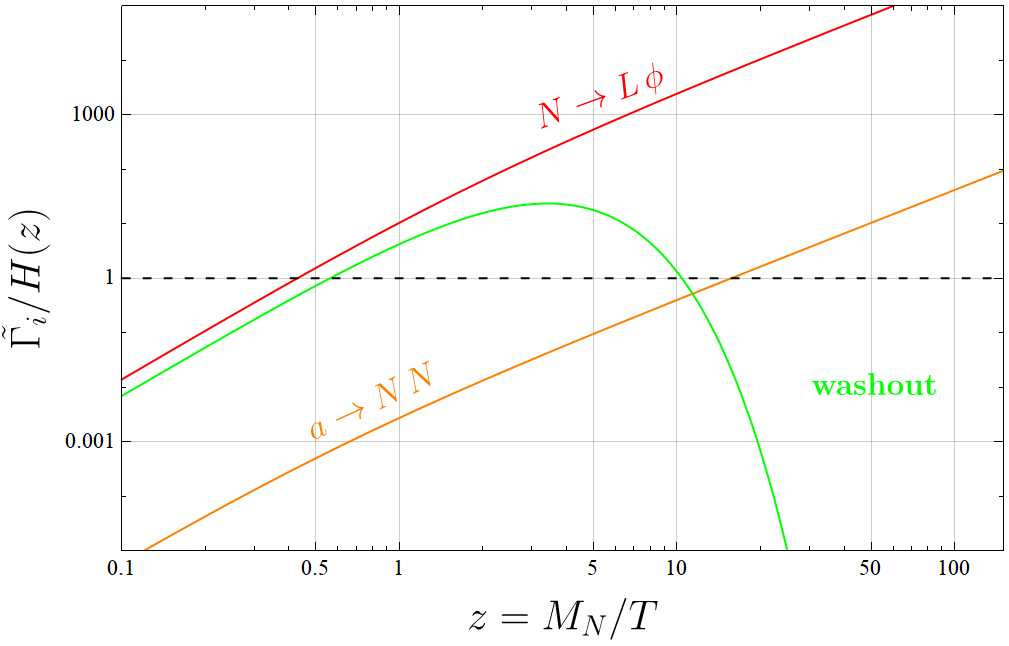}} 
\caption{\small{(Left) Plot of the evolution of the lepton asymmetry Yield (solid blue line) and the RHN Yield (solid black line). The dashed lines show the pure thermal leptogenesis solution. (Right) Rates of the processes involved in ALP leptogenesis, normalized by Hubble parameter. Notice that when the ALP decay becomes efficient ($\tilde{\Gamma} / H >1$), the washout processes are already out-of-equilibrium. Thus, our scenario evades strong washout regime. }}
\label{fig:Boltzmann_solution}
\end{figure}

We present the result of such a procedure in the left-hand panel of Fig.~\ref{fig:Boltzmann_solution}.
In this Figure, we have considered the scenario of resonant leptogenesis with two quasi-degenerate RHN, which are democratically coupled to the ALP, and yield the $CP$ violation $\epsilon = 6 \cdot 10^{-8}$.

In Fig. \ref{fig:Boltzmann_solution} (left), 
the black lines show the evolution of the RHN Yield, while the blue lines the absolute value of $B-L$ Yield. The solid lines represent the total ALP leptogenesis trajectory while the dashed ones are the pure thermal leptogenesis case.
For $z \lesssim 1$, the RHN Yield $Y_N$ increases while RHNs are produced via inverse decay $L \, \phi \to N$ in the thermal bath.
Afterwards, RHNs decay generate a lepton asymmetry $|Y_{B-L}|$.
The sharp bump in $|Y_{B-L}|$ at $z \simeq 1$ is due to the sign flip in the lepton asymmetry. We see now how our new mechanism differentiates from pure standard leptogenesis: the evolution of $Y_N$ in ALP leptogenesis shows a bump at $z \sim 20$, sourced by the non-thermal production of RHNs via ALP's decay. This consequently leads to an increase of the lepton asymmetry, that eventually freezes out at $z \gtrsim 50$. Conversely, in standard thermal leptogenesis, the RHNs quickly drop exponentially for $z \gtrsim 1$, and the lepton asymmetry gets partially washed-out.
Thus, we observe an enhancement of the final asymmetry of around two orders of magnitude. This corresponds to the wash-out factor. This peculiarity of our scenario becomes straightfoward by looking at the right panel of Fig.~\ref{fig:Boltzmann_solution}, where the evolution of the rates $\tilde{\Gamma}_i = \gamma_i/(s Y_{i}^{eq})$ of the RHN decay (red line), the ALP decay (orange line) and the washout processes (green line) (namely RHN inverse decay) are tracked: indeed, when the ALP decay becomes efficient ($\tilde{\Gamma} / H >1$) at $z \simeq 20$, the washout processes are already out-of-equilibrium. Thus, our scenario of ALP leptogenesis evades the strong washout regime, and the lepton asymmetry $|Y_{B-L}|$ gets enhanced with respect to standard leptogenesis at the time of ALP decay, $z \simeq 20$.

\subsection{Results and Discussion}
\label{sec:results_discussion}
In this section, we extend our numerical analysis to the entire parameter space of our interest, and discuss the impact of ALP leptogenesis on the tuning of the mass splitting of RHNs. 

While in the case of thermal leptogenesis we can parameterize the resulting baryon asymmetry as Eq.\eqref{eq:YB},
in the case of ALP leptogenesis, if the ALP decay after the RHN inverse decays have decoupled, i.e. $T^a_D \lesssim M_N/20$, we can approximate the baryon yield by 
\bea 
\label{eq:ALP_lepto}
Y^{\text{ALP lepto}}_{\Delta B}
    \simeq \frac{2Y_a^{\rm initial} \text{Br}_{a \to NN}\,c_{\rm sph}\,\epsilon_{\rm CP}}{D_{\rm{SM}}}\, , 
\eea 
where typically $\text{Br}_{a \to NN} \approx 1$. 
We have checked that this analytical approximation agrees well with the numerical solution in all our parameter space. 
\footnote{
In Eq.\eqref{eq:ALP_lepto}, we consider that the only channel of decay of the RHN is $N \to \phi L$. In principle $\Delta L=1$ scatterings like $tN \to \phi L$, being possibly enhanced by the boost of the $N$, could modify this result. We find in Appendix \ref{app:CP_viol_scat} that Eq.\eqref{eq:ALP_lepto} however remains valid.
}
In general, the final baryon asymmetry in our scenario is given by the sum of Eq.\eqref{eq:ALP_lepto} and Eq.\eqref{eq:YB}, also including the dilution factor for the standard contribution. Note that, if $2 Y_a^{\rm{initial}}>Y_N^{\rm eq} \kappa_{\rm wash} 
$, the 
ALP leptogenesis contribution dominates over the standard one. This will always be the case in our parameter space, so that there Eq.~\eqref{eq:ALP_lepto} is a very good approximation of the final baryon yield.
\\

\begin{figure}[h!]
\centering
\centering
{\includegraphics[width=0.49\textwidth]{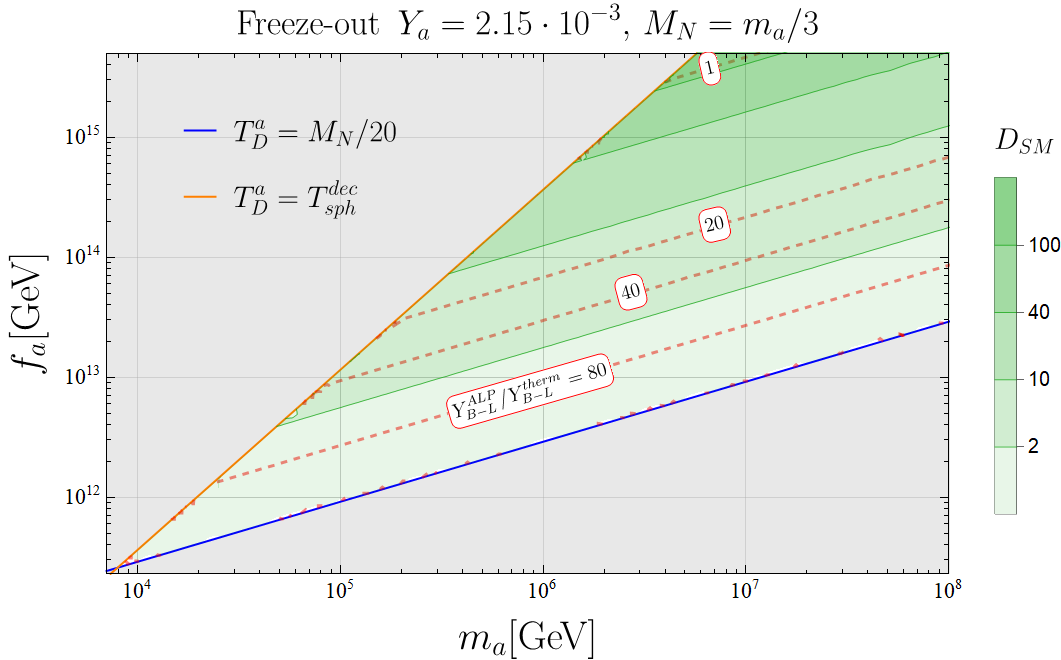}}
\hspace{1mm}
\includegraphics[width=0.49\textwidth]{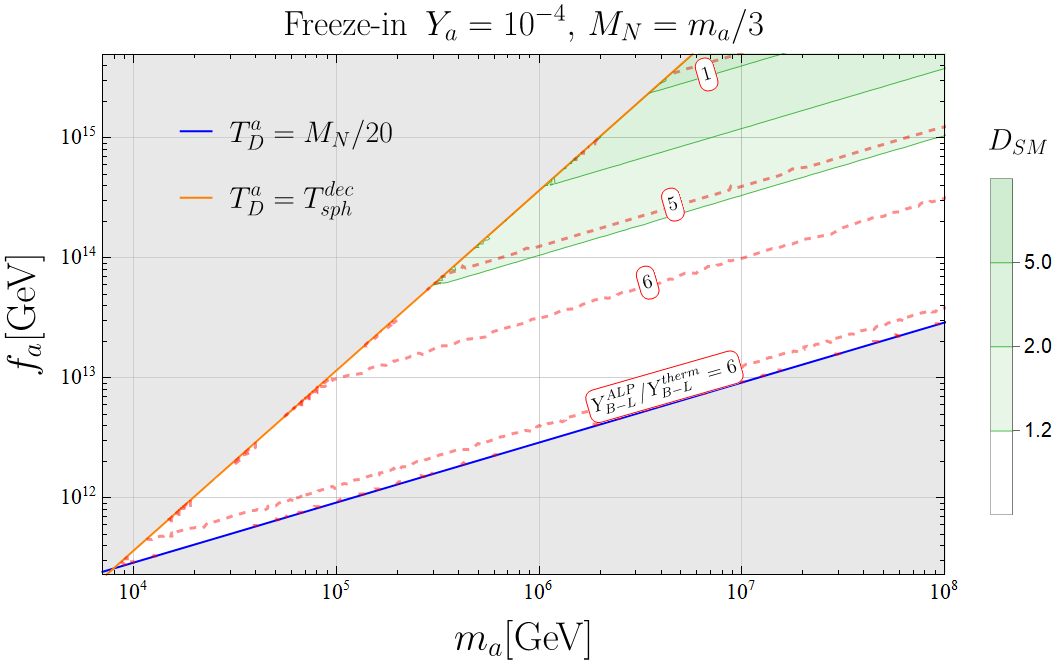}\\
\caption{\small{Contour lines of gain in lepton asymmetry in ALP leptogenesis with respect to the case of thermal leptogenesis in the absence of an ALP. Note that in the latter case dilution is absent, while in ALP leptogenesis the thermal contribution gets also diluted. The green shaded regions show the value of the dilution factor $D_{SM}$.
$M_N= m_a/3, Y_a = 2.15 \times 10^{-3}$ (Left),  $M_N= m_a/3, Y_a =  10^{-4}$ (Right). 
The gain is of course milder when the ALP is frozen-in, because its yield is lower than the one in the freeze-out case.
}}
\label{fig:gain}
\end{figure}

Fig.\ref{fig:gain} 
shows 
the impact of ALP leptogenesis while varying the ALP mass and decay constant.
As previously, the viable region for ALP leptogenesis is the triangle delimited by the 
gray areas.
In this region, we
show contour lines (dashed red lines) of the gain in the final asymmetry in ALP leptogenesis with respect to pure thermal leptogenesis.
The \emph{gain} ratio is defined as 
\bea 
\text{Gain} \equiv \frac{Y^{\rm ALP}_{B-L}}{Y^{\rm \rm therm}_{B-L}}
\eea 
where $Y^{\rm ALP}_{B-L}$ is the yield of ALP leptogenesis computed using Eq.\eqref{eq:ALP_lepto} and $Y^{\rm \rm therm}_{B-L}$ is the ``would-be'' yield from standard thermal leptogenesis with the same see-saw parameters and without the presence of the ALP (and without dilution).
In addition, the value of the dilution factor from the period of matter domination from the ALP is indicated with the green shading. 
We can clearly see that ALP leptogenesis enhances the yield up to two orders of magnitude in the lower part of our viable parameter space (which corresponds to lower values of $f_a$), where the dilution factor $D_{\rm SM}$ is small. Instead, dilution becomes important in the upper part of the plot, weakening the efficiency of our mechanism. 
On the upper part of the plots, dilution is large and there is no gain with respect to standard thermal leptogenesis.

An important implication of our mechanism is that, in the regions where the gain is larger than one, the BAU can be reproduced for RHN masses lower than in standard leptogenesis.
However, the Davidson-Ibarra bound on the mass of hierarchical neutrinos~\cite{Davidson_2002} still implies that the BAU cannot be reproduced if RHNs are lighter than about $10^9$~GeV.

A possibility to realise the observed BAU with lower RHN masses is to rely on resonant enhancements, driven by small mass splittings between the RHNs. Another relevant implication of our mechanism is that the needed tuning, in the RHN mass splittings, is lowered with respect to standard leptogenesis. This tuning can be quantified via \cite{Pilaftsis:1997jf, Pilaftsis_2004}
\bea 
x \equiv \frac{M_1^2- M_2^2}{M_1 M_2} \, ,
\eea 
where we are taking into account only two quasi-degenerate RHNs, $N_1$ and $N_2$ for simplicity. Using this parameter, the $CP$-asymmetry due to decay of $N_1$ takes the following form\cite{Pilaftsis:1997jf}
\begin{equation}
\label{eq:res}
    \epsilon_{1} = \frac{\text{Im} [(y_{\nu}^{\dagger} y_{\nu})^2_{12}]}{8 \pi (y_{\nu}^{\dagger} y_{\nu})_{22}} \frac{x}{x^2 + 4 A_{22}^2}  \, ,
\end{equation}
where $4 A_{22}^2 = \Gamma_{N_2}^2/M_2^2$ is the regulating term. A similar expression for the $CP$-asymmetry parameter $\epsilon_2$ can be found and both the parameters $\epsilon_{1}$ and $\epsilon_{2}$ contribute constructively to $CP$ violation. Here, we assume that the real part of Yukawa coupling has the same order of magnitude of the imaginary part, such that $\frac{Im [(y_{\nu}^{\dagger} y_{\nu})^2_{12}]}{(y_{\nu}^{\dagger} y_{\nu})_{11} (y_{\nu}^{\dagger} y_{\nu})_{22}} \sim 1 $ (and $\Gamma_{N_1} \simeq \Gamma_{N_2}$).  

\begin{figure}[h!]
\centering
\centering
{\includegraphics[width=0.44\textwidth]{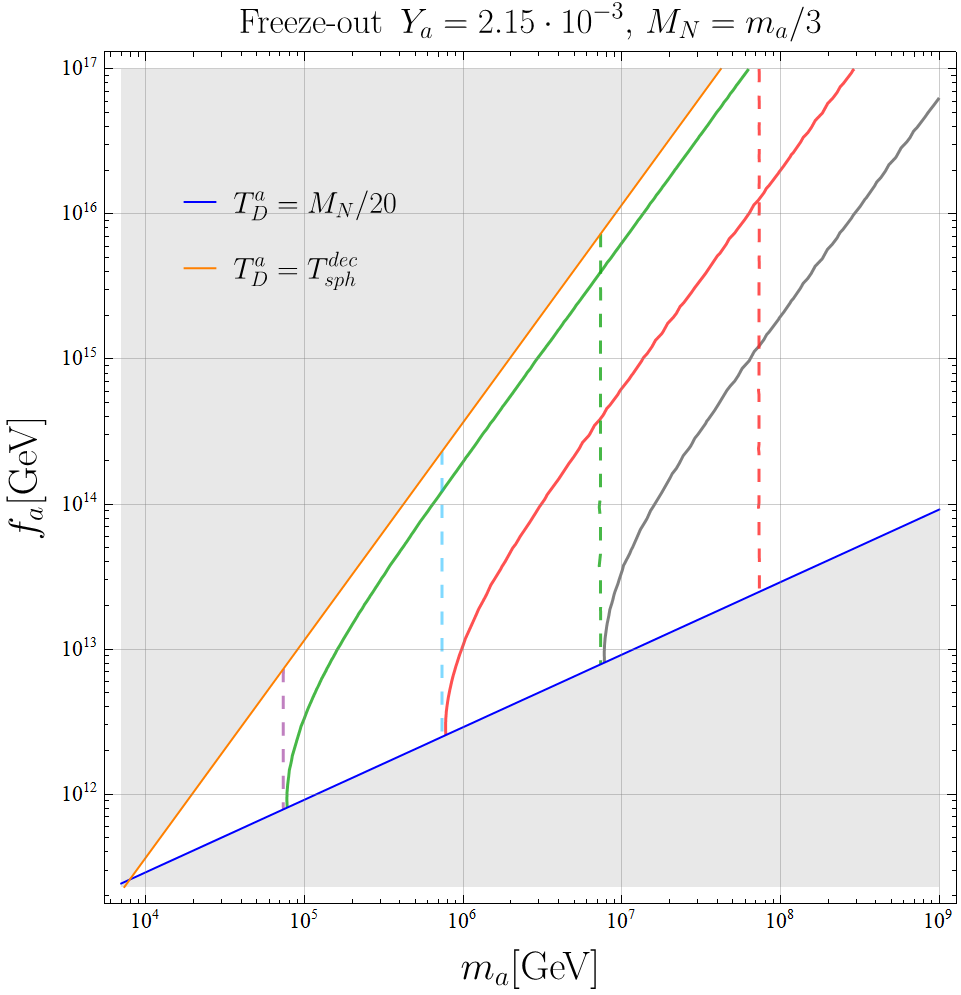}}
\hspace{3mm}
{\includegraphics[width=0.5\textwidth]{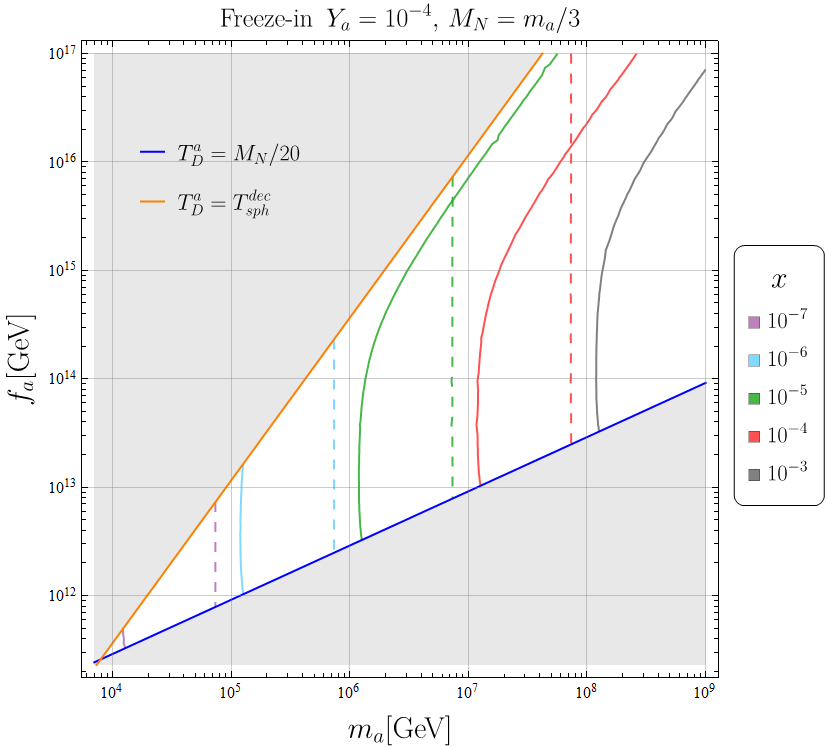}}
\caption{\small{Relative mass splitting $x \equiv (M_1^2- M_2^2)/(M_1 M_2)$ between the RHNs necessary for producing the observed baryon asymmetry in pure thermal leptogenesis (dashed lines) and ALP leptogenesis (full lines).
}}
\label{fig:dilll}
\end{figure}

Let us define $\epsilon_T$ and $\epsilon_{ALP}$ as the $CP$-asymmetry parameters required to obtain the observed value of $Y_{\Delta B}$, respectively in pure standard thermal leptogenesis and ALP leptogenesis.  As we have computed previously, in ALP leptogenesis we gain a factor $\sim \kappa_{\rm wash}^{-1} \sim  100$ in the final $Y_{B-L}$ such that the Majorana mass tuning required is less severe with respect to thermal leptogenesis.  This relaxes the required mass tuning $x_{\rm ALP}$.  In Eq.\eqref{eq:res} we have written the $CP$-parameter as a function of $x$ and we can estimate the gain in mass tuning in ALP leptogenesis as
\begin{equation}
\label{eq:epsres}
\epsilon_{\rm ALP}(x_{\rm ALP})   \simeq \frac{D_{\rm SM}}{2}\frac{Y^{\rm eq}_N}{Y^{\rm initial}_a}\kappa_{\rm wash}\epsilon_{T}(x_T)  \, ,
\end{equation}
where $x_{\rm ALP}$ and $x_{T}$ are the Majorana mass splittings respectively in ALP and thermal leptogenesis. We assumed that the ALP leptogenesis could totally avoid the wash-outs. From Eq.\eqref{eq:epsres}, we obtain the relaxation of the mass tuning in ALP leptogenesis 
\begin{equation}
x_{\rm ALP}  \simeq \frac{2}{D_{\rm SM}}\frac{Y^{\rm initial}_a}{Y^{\rm eq}_N}\kappa^{-1}_{\rm wash} x_T  \, .
\end{equation} 
Therefore, at first approximation, a linear relation between the mass splittings holds (see also \cite{Pilaftsis:1997jf}) and the gain factor is $\sim 10^2$. The relation between $x_T$ and $x_{\rm ALP}$ is studied numerically in Fig.~\ref{fig:dilll} for $ Y_a = 2.15 \times 10^{-3}$ (left panel) and $Y_a =  10^{-4}$ (right panel), when $M_N= m_a/3$.
We observe that the enhancement by two orders of magnitude of $x_{\rm ALP}$ over $x_T$ is realised mostly close to the lower boundary of the triangle region (low $f_a$). At the upper boundary, for large $f_a$, the dilution becomes important and $x_{\rm ALP}$ becomes even smaller than $x_T$.

\section{Susy model realisation and constraints from cosmology}
\label{sec:gravitino}

In the former sections, we presented all the details of a new mechanism of leptogenesis, which we dubbed ALP-leptogenesis, but we did not specified the cosmological constraints that might restrict our scenario. These constraints are unavoidably model dependent.  In this section we fill this gap and present a model to realise our scenario within a SUSY framework, within gauge mediation of SUSY breaking (see e.g. \cite{Martin:1997ns,Giudice:1998ic}). 
In this case the most relevant constraints arise from the cosmology of the gravitino, that we now discuss.

The minimal realisation of ALP leptogenesis only requires the SM to be augmented by the ALP $a$ and the RHN $N_i$, with the minimal couplings as presented in Eq.\eqref{eq:axionl}. In the SUSY implementation, the SM is also augmented by the superpartners, and the ALP is naturally the R-axion, which is the PNGB of the spontaneously broken R-symmetry. 
In addition, the SUSY spectrum includes the gravitino.
ALP leptogenesis will occur in the way we described in the preceding sections, involving the R-axion as the ALP and having the couplings described in Eq.\eqref{eq:axionl}. 

On the other hand, the SUSY 
spectrum now contains several new particles that come with different mass scales.
First, we assume that all the superpartners obtain masses of order one single scale $m_{\rm soft}$ through a gauge mediation mechanism (in order for this to occur for the scalar partner of the RHN, the B-L symmetry should be gauged). 
%
%

Then, 
differently than in non-SUSY scenarios, the ALP mass is not a completely free parameter.
Indeed, within a SUSY realisation, the mass of the R-axion receives an irreducible contribution from the tuning of the cosmological constant, which is given by \cite{Bellazzini:2017neg}
\bea 
\label{eq:irresusy}
\left(
m^{\rm irred}_a \right)^2 \sim \frac{8\sqrt{3} \omega_R F}{M_{\rm Pl} f_a^2} = \frac{24 \omega_R m_{3/2}}{f_a^2}, \qquad \omega_R \equiv c \times \frac{F f_a}{2\sqrt{2}} \, ,
\eea 
where $c$ is a constant smaller than 1 \cite{Dine:2009sw,Bellazzini:2016xrt} and in general small values of $c$ imply tuning in the superpotential. 
We will fix it to $c= 1/2$ in the following for concreteness.

Finally, in our model, like in many SUSY realisations with gauge mediated SUSY breaking, the lightest supersymmetric particle (LSP) is the gravitino $\tilde G$, which is stable with mass
\bea 
m_{3/2} = \frac{F}{\sqrt{3} M_{Pl}} \, .
\eea 
The stability of the Gravitino thus implies that it will constitute a non-zero fraction of the DM abundance of the universe and many different processes of the early universe can produce it copiously. Assuming first that $\tilde G$ is never thermalized in the history of the universe, the gravitino is mostly produced via Freeze-In (FI). Within our model, $\tilde G$ can be frozen in via several processes involving the superpartners like for instance the gluinos $\lambda$. This includes scatterings (UV freeze-in) $g \lambda \to \tilde G $, $g \lambda \to \tilde Gg$ or decays (IR freeze-in) $\lambda \to \tilde G g$,
as well as effective interactions mediated by the superpartners.
We will present now the dominant channels through which the Gravitino can be produced in the early Universe in our set-up.
\begin{itemize}
    \item \textbf{Freeze-in via thermal superpartners}: The gravitino can be produced via the couplings
    to the SM and supersymmetric particles that it acquires through its Goldstino component \cite{Rychkov:2007uq}. As representative interaction we consider the coupling connecting 
    the gluino, the gluons and the goldstino
\bea 
\frac{M_3}{F}\tilde G \sigma_{\mu \nu} \lambda G^{\mu \nu} \, ,
\eea 
where $M_3$ is the mass of the gluino. The gravitino can be produced if an abundance of gluino occurs in the bath via i) scatterings $\lambda g \to g \tilde G$, ii) directly via decay of the gluino $\lambda \to  g \tilde G$.
Similar processes exist also for the other superpartners in the thermal bath (including possibly the sneutrinos). In addition, the gravitino can be produced via the late decay of the lightest observable supersymmetric particle, typically the Bino, with mass $M_1$.
All these processes bring unacceptably large abundance of the gravitino. This is a manifestation of the gravitino problem, see e.g.~\cite{Moroi:1993mb,Kawasaki:1994af,Kawasaki:2008qe}. To avoid this issue, we assume that the
soft masses $m_{\rm soft}$
of all sparticles, generated through gauge mediation, are 
much larger than the reheating temperature: 
\bea
T_{\rm RH} \ll m_{\rm soft} \sim M_1 \sim M_3 \sim \frac{\alpha}{4\pi} \frac{F}{M_{\rm mess}} \qquad \text{(NO gravitino FI from sparticles)} \,. 
\eea 

In such regime, the superpartners 
are never produced copiously in the bath and the production of the Gravitino via those channels is suppressed.

\item {\textbf{ Freeze-in via SM particles}} The gravitino can be produced via FI by 
effective interactions involving the SM fields which are generated by integrating out the heavy superpartners. 
We can estimate the order of magnitude of these contributions as follows.
Integrating out the
superpartners one typically 
obtains effective operators involving two Goldstino suppressed by two powers of $F$ 
(see e.g. \cite{Brignole:1997sk}).
The corresponding yield can be 
estimated as:

\bea 
Y_{\tilde G}^{\rm{High-Dim}} 
\sim 
\left(\frac{T_{\rm RH}}{F}\right)^4 T_{\rm RH}^3
M_{\rm Pl}
\eea

which will find to be always subdominant in our parameter space. We will thus ignore it in the rest of this analysis. 

\item \textbf{Freeze-in via the decay of the R-axion}: An unavoidable contribution to the abundance of gravitino originates from the R-axion decay.
This turns out to be the most relevant channel within our assumptions, eventually determining the gravitino abundance.
The decay rate $a \to \tilde G \tilde G  $ is given by \cite{Bellazzini:2017neg}
 \bea 
 \label{eq:decay_rate}
 \Gamma(a \to \tilde G \tilde G) \approx \frac{1}{4\pi} \frac{m_a^5 \omega_R^2}{f_a^2 F^4} \,. 
 \eea 
Assuming that the other gravitino production mechanisms discussed previously are negligible, 
the DM yield is then given by 
 \bea 
 Y_{\tilde G} \approx 2 Y_a \text{Br}[a \to \tilde G \tilde G], \qquad \text{Br}[a \to \tilde G \tilde G] \equiv \frac{ \Gamma(a \to \tilde G \tilde G)}{\Gamma(a \to \text{All others})+  \Gamma(a \to \tilde G \tilde G)} \, ,
 \eea 
Using the expressions in Eq.\eqref{eq:decay_rate} and in Table \ref{tab:channels},
considering that the dominant decay channel for the R-axion is into RHN, the yield of the gravitino becomes
\bea 
 Y_{\tilde G} \approx 4 Y_a \frac{m_a^4\omega_R^2}{M_N^2F^4 \sqrt{1-4 \frac{M_{N}^2}{m_{a}^2}}}  \, .
 \eea 
 Taking $\omega_R $ as in Eq.\eqref{eq:irresusy} and $M_N \sim m_a/3$,
 we get
\bea 
 Y_{\tilde G} \approx \frac{36}{\sqrt{5}}Y_a\frac{m_a^2 c^2 f_a^2}{ m_{3/2}^2 M_{\rm Pl}^2 } \,,
\eea 
where we used $m_{3/2} = F/(\sqrt{3} M_{\rm Pl})$.

This leads to the following abundance of gravitino today
 \bea
 \label{eq:yield_DM}
 \Omega_{3/2} h^2  = \frac{m_{3/2} Y_{\tilde G} s_0}{\rho_c/h^2} \approx 2.2 \times 10^8 \bigg(\frac{m_{3/2}}{\text{GeV}} \bigg) Y_{\tilde G}
 \eea 
 where $\rho_c$ is the critical energy density today and $s_0$ is the entropy density today. We need to require that the DM density in Eq.\eqref{eq:yield_DM} is smaller than $ \Omega^{\rm DM}_{\rm obs} h^2 = 0.12$~\cite{Planck:2018vyg}.  

\end{itemize}

\noindent
To summarize, the previous considerations require the following hierarchies
\bea 
\label{eq:hierarchies}
 m_{\rm soft} \gg T_{\rm RH} > m_a > m_{3/2} \,.
\eea 
In this scenario, the aftermath of SUSY consists into a moderately heavy R-axion and an LSP gravitino, produced through the R-axion decay. 
Because of the assumptions about the soft mass scales and $T_{\rm RH}$, the R-axion is produced in the Early Universe only through the couplings with the SM as in \eqref{eq:axionl} and as discussed in Sec. \ref{sec:Alp_prod}.

The abundance of the R-axion determines the effectiveness of the
leptogenesis as well as the dark matter (gravitino) abundance.
Because of the low reheating temperature requirement, the natural setting is the freeze-in production for the ALP.

\begin{figure}[h!]
\centering
\includegraphics[width=9cm]{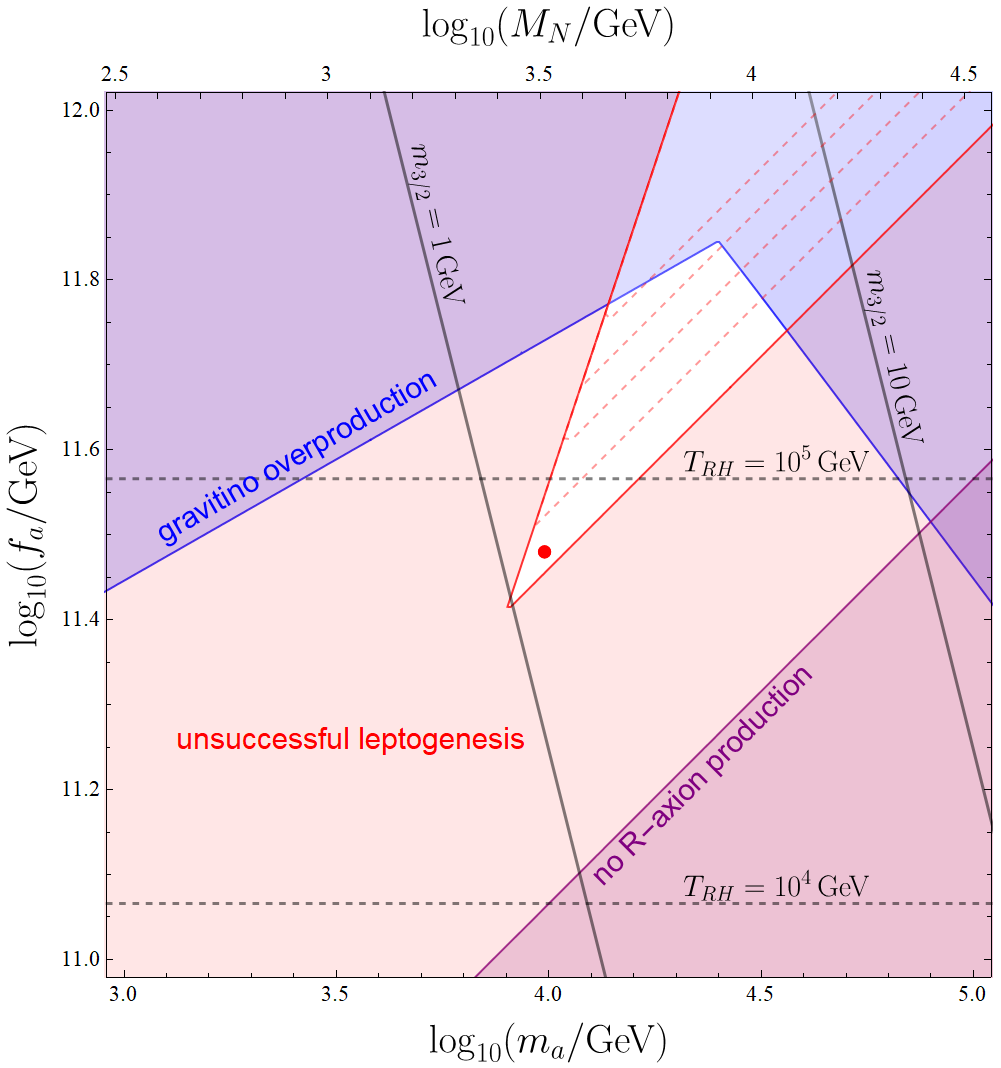}
\caption{\small{Viable parameter space for SUSY realisation with an $R$-axion, with ALP freeze-in abundance $Y_a = 10^{-4}$ and RHN mass $M_N = m_a/(3,3.5, 4,4.5, 5)$, corresponding to red dashed lines. The shaded regions represents cosmological bounds. The red area corresponds to unsuccessful leptogenesis. The white region accommodates both baryogenesis and gravitino production. Slightly smaller ratios of $M_N/m_a$ modify the viable ALP leptogenesis region, closing the white region, as shown by the red dashed lines. Successful leptogenesis predicts $m_{3/2} \simeq 1$ GeV, with $m_a \sim 10$ TeV and $f_a \sim 5 \times 10^{11}$ GeV (denoted by the red point).
}
}
 \label{fig:cosmo_constraints}
\end{figure}



In Figure \ref{fig:cosmo_constraints}, we present the constraints on the specific realisation of this model.
We consider the $R$-axion to be frozen-in with abundance $Y_a = 10^{-4}$, which fixes the reheating temperature as a function of $f_a$, see Figs.~\ref{fig:freezeoutin} and~\ref{fig:dil1}.
The SUSY breaking scale $F$ (and hence the gravitino mass) is determined by requiring that the R-axion mass is given by the irreducible contribution in \eqref{eq:irresusy}.
This fixes all the parameters of the model as a function of $f_a$ and $m_a$ and allows us to impose the various constraints.

The region displaying gravitino overproduction is shaded in blue.
The upper left part is excluded
because the reheating temperature is too small to satisfy the requirement in Eq.~\eqref{eq:hierarchies} for the soft masses.
In order to impose this, we estimated 
$m_{\rm soft} \approx \frac{\alpha_s}{4 \pi} \sqrt{F}$ and we required $T_{\rm RH}/m_{\rm soft} < 10^{-2}$.
The right upper part is excluded because the gravitino production through R-axion decay is too large (see Eq.~\eqref{eq:yield_DM}).
In the region shaded in purple 
the requirement of the freeze-in 
R-axion abundance would require the reheating temperature to be smaller than the R-axion mass, so it is not consistent.
On the top of this, the light-red region excludes parameter space where ALP leptogenesis is not successful,
getting a viable spike-like shape as already identified in Figures \ref{fig:dil1} and \ref{fig:gain} (and no matter domination occurs here).

Remarkably, the combination of all those requirements selects a small white region that can induce ALP leptogenesis with values: $m_a \sim 10$ TeV, $f_a \sim 5 \times 10^{11}$ GeV and $m_{3/2} \sim 1$ GeV.
In the bulk of this region the gravitino is under abundant, while it constitutes the entirety of the dark matter on the blue line.

While we fixed $Y_a = 10^{-4}$ for this plot, 
increasing the yield $Y_a$ closes the white region due to stronger overproduction of gravitino DM.
Reducing the model-dependent value of $\omega_R$, instead, enlarges the viable portion of parameter space since it reduces the BR of $a$ into gravitinos.
In the plot, we also showed the effect of slightly changing the ratio $M_N/m_a$, whose only effect is to modify the viable ALP leptogenesis region (this is indicated in the plot with dashed red-lines).
Besides order one numbers, our analysis shows that the embedding of ALP leptogenesis in SUSY models can lead to very predictive scenarios 
for the cosmological history, with relatively low-scale BSM particles, possibly targets for future colliders.

In Appendix \ref{app:SUSY_model}, as a minimal existence proof, we provide a concrete realisation of a SUSY model displaying a low energy spectrum which is populated by the SM, the R-axion, the heavy neutrinos, and the gravitino, while all the superpartners, including the sneutrinos lie at much larger energy scale, $M_1  \sim M_3 \sim m_{\rm soft}$.

\section{Summary and outlook}
\label{sec:summary}

In this article we have proposed a new mechanism of non-thermal leptogenesis where an axion-like particle (ALP), which can initially be at equilibrium or not with the early universe bath, decays dominantly to right-handed neutrinos (RHNs) themselves generating a lepton asymmetry upon decaying. A sketch of our mechanism and of its cosmological stages is displayed in Fig.~\ref{fig:schematic}, and the final yield for the baryon asymmetry of the universe is given as a simple expression in Eq.~\eqref{eq:ALP_lepto}.

Since the RHNs are produced out of equilibrium they do not undergo the usual thermal wash-out effects, and leptogenesis is enhanced by up to two orders of magnitude with respect to standard thermal leptogenesis. We have identified the entire parameter space where ALP leptogenesis improves over standard one, see Fig.~\ref{fig:gain}. This includes a region where the ALP dominates the energy density of the universe before decaying, and so injects entropy in the bath and dilutes relics, including itself and so the lepton asymmetry. In that region we have included such dilution in our BAU computation.

The gain of ALP leptogenesis with respect to standard one translates either in a lower RHN mass capable to reproduce the BAU and neutrino masses, or in a milder tuning of the RHN mass splitting if one goes in the regime of resonant leptogenesis, needed e.g. for RHNs at the TeV scale. The latter tuning is visualized in Fig.~\ref{fig:dilll}.

    We finally embed our scenario in a SUSY completion (having in mind gauge mediation of SUSY breaking) where the ALP is the R-axion (the PNGB of the R-symmetry breaking) and the gravitino is the natural DM candidate. This model turns out to be surprisingly predictive. The model selects a narrow range of parameters where $M_N \sim m_a/3$ with $m_a \sim 10^4$ GeV,
    $f_a \sim 5 \times 10^{11}$ GeV,
    and the gravitino with mass $m_{3/2} \sim 1$ GeV (i.e. $\sqrt{F} \sim 2 \times 10^{9}$ GeV). In this narrow window, visualized in Fig.~\ref{fig:cosmo_constraints}, the correct lepton asymmetry is obtained, and the gravitino is the DM. 

There are several possible directions on which our work could be extended. 
In the see-saw scenario we consider, the mixing angle with SM neutrinos would be too small to potentially lead to interesting signatures. 
However, in the case of an inverse see-saw~\cite{Mohapatra:1986aw,Mohapatra:1986bd} realisation of ALP leptogenesis, one might increase the mixing angles and renders the RHN accessible in a future high-energy muon collider, see e.g.~\cite{Li:2023tbx,Dichtl:2023xqd}. Note that a larger mixing angle could however also impact production rates in the early Universe.

Then, in terms of UV completions of our scenario,
one could explore the possibility 
of identifying the ALP with the Majoron \cite{Chikashige:1980qk,Chikashige:1980ui,Gelmini:1980re}. While the majoron couples naturally with RHN and leptons, 
it does not with gluons and quarks, and hence there will be modifications in its production rates in the early Universe. In addition, the necessary sizeable mass for the ALP implies in this case a sizeable explicit breaking of the lepton symmetry, possibly leading to complementary phenomenological signatures.
We leave to the future the detailed investigation of such interesting directions in ALP leptogenesis.\\

\emph{Note:}  While this paper was typewritten, another paper discussing the enhancement of leptogenesis via the decay of BSM scalar fields~\cite{Tong:2024lmi} appeared.
Differences include that they focused on CP-even scalars while we focused on CP-odd ones, that we discuss an explicit UV realisation that predicts a sharp connection of the mechanism with dark matter, and that our cosmological analysis extends to the regimes i) of freeze-in of the scalar and not only freeze-out, ii) where the scalar predicts early matter domination in the early universe, affecting leptogenesis.

\section*{Acknowledgements}
We thank Silvia Pascoli for collaboration in the early stages of this work, and Alessandro Granelli for useful comments. FS thanks Yue Zhang for useful discussions on dilution at Portoroz 2023.

{\small
AM and MV are supported by the ``Excellence of Science - EOS" - be.h project n.30820817, and by the Strategic Research Program High-Energy Physics of the Vrije Universiteit Brussel,
and by the iBOF ``Unlocking the Dark Universe with Gravitational Wave Observations: from Quantum Optics to Quantum Gravity'' of the Vlaamse Interuniversitaire Raad. 
MC is supported by the Deutsche Forschungsgemeinschaft under Germany´s Excellence Strategy - EXC 2121 "Quantum Universe" - 390833306.
FS is supported in part by the European Union's Horizon research and innovation programme under the Marie Sklodowska-Curie grant agreements No.~860881 - HIDDeN and No.~101086085 - ASYMMETRY, by COST (European Cooperation in Science and Technology) via the COST Action COSMIC WISPers CA21106, and by the Italian INFN program on Theoretical Astroparticle Physics.}

\appendix

    \section{SUSY realisation of ALP leptogenesis}
\label{app:SUSY_model}

We have seen in the main text that having superpartners with masses close to the ones of their counterparts would unavoidably lead to the overproduction of gravitino when the superpartners decay. In this appendix, we propose a SUSY realisation where the superpartners (including the sneutrinos) all acquire a large mass and are never produced by reheating. 
This can be obtained by gauging the $U_{B-L}(1)$ symmetry
(see for instance~\cite{Babu:2009pi} for similar constructions).
In addition, to obtain the required R-axion coupling to the RHNs, the R-charge of the RHN supermultiplet should be such that its mass is protected by the R-symmetry.

To achieve this,
we introduce a sector with the following superfields and charges. Note that this model is developed for one single RHN. It can be easily generalized to more RHN's, which is necessary in order to gauge the B-L symmetry.
The new sector includes, on the top of the superfield $N$ containing the neutrinos and the sneutrino, five new superfields: $Y$, $\tilde Y$, $R$, $\tilde R$, which will be charged under $B-L$, and a singlet $S$. Furthermore, these fields interact also with $\Phi$, the superfield 
carrying the SUSY breaking $F$ term and the R-symmetry breaking vev $f_a$,
induced by a SUSY (and R-) breaking sector.
The superfield $\Phi$ contains 
in its fermionic component
the goldstino, while the scalar component includes as real part the pseudomodulus (assumed to be decoupled and with mass $\sim \sqrt{F}$) and as phase the R-axion.
In terms of constrained superfields, this would correspond to the constrained chiral superfield containing both the goldstino and the R-axion, denoted as $\Phi_{NL}$ in Appendix E of \cite{Komargodski:2009rz}.
The $R$ and $B-L$ assignments of all those field are presented in Table \ref{tab:cross_sec}.

\begin{table}[t]
\begin{center}
\begin{tabular}{cccccccc}
\hline
\multicolumn{1}{|l|}{Superfield} & \multicolumn{1}{l|}{$\Phi$} 
& \multicolumn{1}{l|}{$Y$}
& \multicolumn{1}{l|}{$\tilde Y$}
& \multicolumn{1}{l|}{$R$}
& \multicolumn{1}{l|}{$\tilde R$}
& \multicolumn{1}{l|}{$S$}
& \multicolumn{1}{l|}{$N$} \\
\hline
\multicolumn{1}{|l|}{$B-L$ number} 
& \multicolumn{1}{l|}{$0$}
& \multicolumn{1}{l|}{$-2$}
& \multicolumn{1}{l|}{$2$}
& \multicolumn{1}{l|}{$-2$}
& \multicolumn{1}{l|}{$2$}
& \multicolumn{1}{l|}{$0$}
& \multicolumn{1}{l|}{$-1$}
\\
\multicolumn{1}{|l|}{$R$ number}
& \multicolumn{1}{l|}{$2$} & \multicolumn{1}{l|}{$0$}
& \multicolumn{1}{l|}{$0$}
& \multicolumn{1}{l|}{$0$}
& \multicolumn{1}{l|}{$2$}
& \multicolumn{1}{l|}{$2$}
& \multicolumn{1}{l|}{$0$}
\\
\hline
\end{tabular}
\end{center}
\caption{$R$ and $B-L$ assignments of each new superfield.
} 
\label{tab:cross_sec}
\end{table}

In the UV, we consider the following superpotential 
\footnote{The superpotential 
in Eq.~(\ref{eq:superpot})
is not the most general one
compatible with the charges in Table~\ref{tab:cross_sec}. 
We assume that other possible terms (like $Y \tilde{R}$) are suppressed in the UV by further symmetries
(this is then preserved at other scales thanks to $W$ non-renormalization).
}
\bea 
W = \bigg(Y \tilde Y - \mu^2 \bigg) S + M R \tilde R - \tilde R N N - R \Phi \tilde Y \, ,
\label{eq:superpot}
\eea 
with the following hierarchy of scales
\bea 
\label{eq:scale_hierarchy}
M \gg \mu \gg  f_a \gtrsim \sqrt{F} \, .
\eea
For energies $M \gg E \gg \mu $, 
we can integrate out supersymmetrically the fields $R$ and $\bar R$ by solving their F-term equations, and the resulting superpotential takes the form 
\bea 
W = \bigg(Y \tilde Y - \mu^2 \bigg) S - \frac{\Phi \tilde Y NN}{M} \, .
\eea 
 Since $Y, \tilde Y$ are charged under $B-L$ which is a gauged symmetry, the potential also receives  D-terms contributions of the form 
\bea 
V_D \propto g_{B-L}^2 \bigg( |Y|^2 -  |\tilde Y|^2\bigg)^2 \, .
\eea 
For energies $E \ll \mu$, the vacuum  takes the following form 
\bea 
Y = \tilde Y = \mu,  \quad S = 0 \, .
\eea 
This breaks spontaneously the $B-L$ gauge symmetry supersymmetrically,
giving a mass of order $\mu$ to the $B-L$ gauge bosons supermultiplet. 
Non-zero $R$-axion couplings to gluons and tops, as assumed in our phenomenological analysis in Sec.~\ref{sec:Alp_prod}, can be generated without Yukawas carrying $R$-charge by gluino loops (whose Majorana mass breaks $U(1)_R$) and via a mixing between the $R$-axion and the MSSM CP-odd Higgs, see e.g.~\cite{Bellazzini:2017neg}.
Notice that, in this case, the $R$-charge assignment for the RHNs in Table~\ref{tab:cross_sec} is not compatible with $SO(10)$ unification  but  it is with $SU(5)$ unification.

The breaking of SUSY and of R-symmetry is assumed to be induced on the field $\Phi$ by a separate sector, implying that the field 
$\Phi$ takes a vev $f_a$ and a non-vanishing F-term $F$.
As a result, the neutrino gets a supersymmetric mass in the effective superpotential as
\bea 
W \supset
\frac{\mu f_a}{M} NN \, , 
\eea 
which gives mass to the fermionic component.
The same superpotential term 
includes also the R-axion coupling to the RHN, proportional to the mass.
Let us notice that in order to map this interaction to the Lagrangian in \eqref{eq:axionl} one has to perform an axion dependent rotation on the RHN, moving the ALP coupling from the mass term to the derivative interaction. This would also induce a non-derivative coupling of the RHN with the ALP in the Yuwaka coupling of the RHN. We do not speficy the origin of the RHN Yukawa interaction here, where possibly other powers of $R$-axion insertion could appear. We notice that in any case the resulting non-derivative coupling would be suppressed by the smallness of the RHN Yukawa interaction.
The 
scalar components of the neutrino superfield are split because of the SUSY breaking contribution as 
\begin{equation}
    m_{\tilde \nu_R}^2 = 
\left( 
\frac{\mu f_a}{M} \right)^2 
\pm \frac{\mu}{M} F 
\sim 
\left(
\frac{\mu f_a}{M} \right)^2
\end{equation}
where in the last
step we assumed 
$f_a^2 \gtrsim F$.

On the top of this contribution,
we assume that the SUSY breaking sector includes messengers (of mass $\simeq \sqrt{F}$) charged under the SM gauge group as well as under the B-L symmetry.
The SM superpartners will acquire soft masses of order $m_{\rm{soft}} \sim \frac{\alpha}{4\pi} \sqrt{F}$.
The
sneutrino will get a gauge mediated SUSY breaking soft mass of the order
\bea 
m^2_{\rm soft} \approx \left( \frac{\alpha_{B-L}}{4\pi} \right)^2
\frac{F^2}{\mu^2} \, ,
\eea 
where the scaling of the gauge mediated contribution of a massive
gauge boson has been taken from
\cite{Gorbatov:2008qa},
in the regime $\mu > \sqrt{F}$.

We then conclude that
 the mass of the neutrinos and the sneutrinos is given by 
\bea 
m_{\nu_R} = \frac{\mu f_a}{M}, \qquad  m_{\tilde \nu_R}^2 \approx \left( \frac{\mu f_a}{M}
\right)^2+ \left( \frac{\alpha_{B-L}}{4\pi}\frac{F}{\mu} \right)^2 \, .
\eea 
These scalings should be consistent with: i) $m_{\tilde \nu_R} \gg m_{\nu_R}$ so that the sneutrinos do not play any role in gravitino production; ii) $m_{\nu_R}$ should be of the same order of the axion mass.
These requirements, 
employing the scaling of the irreducible R-axion mass in Eq.\eqref{eq:irresusy},
$m_a^2 \sim F^2/ (f_a M_{\rm Pl})$,
transfer into the following statements on the energy scales of our model 
\bea 
\frac{F}{f_a^{1/2}M_{\rm Pl}^{1/2}} \sim \frac{\mu}{M}, \qquad \frac{\alpha_{B-L}}{4\pi}\frac{f_a}{\mu}  \gg \frac{f_a^{1/2}}{M_{\rm Pl}^{1/2}} \, .
\eea 
which is consistent with the requirements that $M < M_{\rm Pl}$, $\mu > f_a$. 
This concludes this appendix, which should be considered as an existence proof of a UV SUSY model realising our ALP leptogenesis scenario.

\section{The origin of $CP$ violation}
\label{app:CP_viol_scat}
In this appendix, we discuss the origin of the $CP$ violation and we comment on the impact from the $\Delta L =1$ scatterings of the type $N t \to Q_3 L$.

After being produced by the ALP decay, RHNs are possibly  strongly boosted  with $E_N \sim m_a/2$. We now analyse the fate of the those RHNs. The two dominating reaction rates involving the boosted RHN are: i) the \emph{decay}, which rate in the plasma frame reads
\bea 
\label{eq:decay_rate}
\Gamma_{N \to \phi L} =\frac{y_{\nu}^2 M_N}{8 \pi \gamma_E} \simeq \frac{y_{\nu}^2 M_N^2}{4 \pi m_a}  \, ,
\eea 
and ii) the $CP$-violating $2 \to 2$ $\phi$-mediated scattering $Nt \to Q_3 L$ with rate\cite{ Giudice_2004, Pilaftsis_2004} (neglecting thermal masses and subleading pieces) 
\begin{align}
\label{eq:rate_scatterings}
\sigma_{Nt \to Q_3 L} = \frac{3 y_\nu^2 y_t^2}{4\pi M_N^2 x} \bigg[ \bigg(\frac{x-1}{x}\bigg)+ \text{logs}\bigg]\qquad \Rightarrow \Gamma_{Nt \to Q_3 L} \approx  n_t \sigma_{Nt \to Q_3 L}  \, ,
\end{align}
where $n_t = g_t \zeta(3) T^3/\pi^2$ with $g_t$ is the number of dof of the tops in the plasma, $y_t$ the top Yukawa coupling and $x \equiv s/M_N^2$ where $s$ is the usual Mandelstam variable. 
 
We now compute the dominant piece of the $CP$ violation from the scatterings. In the frame of the boosted $N_1$, we obtain the following factorisation 
\bea 
(\sigma(N t \to Q_3L) - \bar{\sigma}(N t \to Q_3L)) (\text{wave})  \approx \underbrace{ f^{\phi L, loop}_{\text{wave}} \times \frac{M_1M_2}{M_1^2 - M_2^2} }_{= \text{CP-violation $N \to \phi L$(\text{wave})}} \times \sigma^{\text{CP conserving}}_{N t \to Q_3L} \, . 
\eea 

In a similar way, the wave part of the decay amplitude could be factorised into 
\bea 
(\Gamma_{N \to \phi L} - \bar\Gamma_{N \to \phi L}) (\text{wave}) = f^{\phi L, loop}_{\text{wave}} \times \frac{M_1M_2}{M_1^2 - M_2^2} \times \Gamma^{\text{tree level}}_{N \to \phi L} \, . 
\eea 
From that we can conclude that 
\bea 
\label{eq:ratio of epsilons}
\epsilon_{N t \to Q_3L}(\text{wave}) =   \epsilon_{N \to \phi L}(\text{wave}) \, ,
\eea 
in agreement with~\cite{Pilaftsis:2003gt, Nardi:2007jp}. 
\emph{This factorisation pattern only holds for the wave piece of the amplitude}.  We can also observe that in general 
\bea 
\epsilon_{N t \to Q_3 L}(\text{vertex}) \neq   \epsilon_{N \to \phi L}(\text{vertex}) . 
\eea 
However, in the resonant regime,  only the wave piece is resonantly enhanced, and it is expected that $\epsilon(\text{wave}) \gg \epsilon(\text{vertex}) $ and the vertex contribution can be neglected altogether. As a consequence, the lepton number induced would take the form 
\bea 
Y_{\Delta L} \approx Y_N (\epsilon_{N t \to Q_3 L} \kappa_{N t \to Q_3 L} f_{N t \to Q_3 L} + \epsilon_{N \to \phi L} \kappa_{N  \to \phi L} f_{N  \to \phi L} )
\eea 
where $\kappa$ represent the wash-out factors that we can safely approximate to be $\approx 1$  and $f_{N t \to Q_3 L}$ ($f_{N  \to \phi L} $) is the fraction of $N$ decaying via scattering (usual decay) such that $f_{N t \to Q_3 L}+ f_{N  \to \phi L} \approx 1 $, assuming that those are the only two channels. 
On the other hand, $\epsilon_{N t \to Q_3 L} (\epsilon_{N  \to \phi L})$ is the $CP$ violation in scattering (decay).
So that 
\bea 
\label{eq:Y_DeltaL}
Y_{\Delta L} \approx Y_N \epsilon_{N \to \phi L}( 1 - f_{N  \to \phi L} +  f_{N  \to \phi L} ) = Y_N \epsilon_{N \to \phi L} . 
\eea
Eq.~\eqref{eq:Y_DeltaL} shows that the channel of decay of $N$ does not have any impact on the final lepton asymmetry. The consequence of this pattern is that the Lepton number left from a fixed number of initial $a$, $n_a \approx n_N/2$, is independent on the channel through which it goes to the standard model (via decay or scatterings).

To get an order of magnitude estimate of the rate of the scattering, we compute the value of $x$ by
\bea 
p_{\rm RHN} = (\sqrt{M_N^2 + p_i^2}, 0, 0, p_i), \qquad p_i \approx m_a/2 ,
\qquad 
p_{\rm t} = (T, 0, 0, -T)
\nn 
\Rightarrow \qquad  s = (p_{\rm RHN} + p_{\rm t})^2 \approx M_N^2 + 4 p_i T\approx M_N^2 + 2m_a T \qquad \Rightarrow \qquad x \approx 1 + 2\frac{m_a T}{M_N^2} \, , 
\eea 
for $p_i \gg M_N$. Comparing the rate of the scatterings in Eq.\eqref{eq:rate_scatterings} with the rate of the RHN decay in Eq.\eqref{eq:decay_rate}, we observe that, for low temperatures $T \lesssim M_N/20$ relevant for ALP leptogenesis, the scatterings are always strongly subdominant with respect to the decays, and we can conclude that scatterings are irrelevant for ALP leptogenesis.

\section{Numerical study of the ALP dilution}
\label{app:dilution}

We can define the dilution factor as the ratio of comoving entropy $S= s a^3$ after and before the decays 
\begin{equation}
\label{eq:DSMentropyratio}
    D_{\rm SM}  \equiv \frac{S_{\rm SM}^{\rm after}}{S_{\rm SM}^{\rm before}}  \, ,\qquad 
    D_{\rm SM}  = \bigg[1 + \bigg(0.43\times \frac{Y_a}{Y_a^{\rm FO}}g_{\star}^{1/4}\frac{g_{a}}{g_{*}} \frac{m_{a}}{\sqrt{\Gamma_{a} M_{\rm Pl}}} \bigg)^{4/3} \bigg]^{3/4} \,,
\end{equation}
where $g_{a} = 1$, $Y_a$ is the ALP yield and $Y_a^{\rm FO}  \simeq 2.15 \times 10^{-3}$ is the freeze-out abundance of the ALP.
 
 In this appendix, we study this dilution numerically. We implement the following system of BEs that tracks the evolution of the energy densities of the axion, the RHN and the radiation and the $B-L$ asymmetry:
\begin{equation}
  \label{eq:dilsyst}
  \text{Total BEs system:} \qquad
    \begin{cases}
    \dot{ \rho}_a = -3 \rho_{a}  - \frac{\Gamma_a}{H} \rho_a \\

    \dot{\rho}_N = -  3 \rho_N + \frac{\text{Br}_{N} \Gamma_a}{H}\rho_a - \frac{\Gamma_D}{H} \rho_N\\
  \dot{\rho}_R = - 4 \rho_R + \frac{(\text{Br}_{g}+\text{Br}_{t}) \Gamma_a}{H} \rho_a +\frac{\Gamma_D}{H} \rho_N \\
  \dot{ Y}_{B-L}/a_0=  - 3  Y_{B-L}\Big(1 + \frac{\dot{\rho}_R }{4 \rho_R }\Big) +\frac{\gamma_{D} \epsilon_{\rm CP}}{H s} \Big(\frac{Y_{N}}{Y_{N} ^{\rm eq}}-1\Big) -  \frac{\gamma_D}{H s} \frac{Y_{B-L}}{2 Y_{l}^{\rm eq}}
    \end{cases}
     \, ,
 \end{equation}
 where the derivatives are computed with respect to the natural logarithm of the scale factor $a$, $\dot{x}= \frac{d x}{d\rm ln (a/a_0)}$, and the factor $-4 \rho_R$ ($-3 \rho_{a, N}$) accounts for the scaling behaviour of radiation (matter). $a_0$ is a normalisation that we set initially. $\text{Br}_i$ denotes the total branching ratio to the decay product $i$. The Hubble parameter is $H = \frac{1}{M_{Pl}} \sqrt{8 \pi( \rho_a +\rho_N +\rho_R)/3}$. We take into account the Fermi-Dirac statistics to write the equilibrium number density for the RHN as $Y_N^{\rm eq} = \frac{3}{4} \zeta(3) \frac{45}{\pi^4 g_*}$.

\begin{figure}[h!]
\centering
\subfigure
{\includegraphics[width=8.7cm]{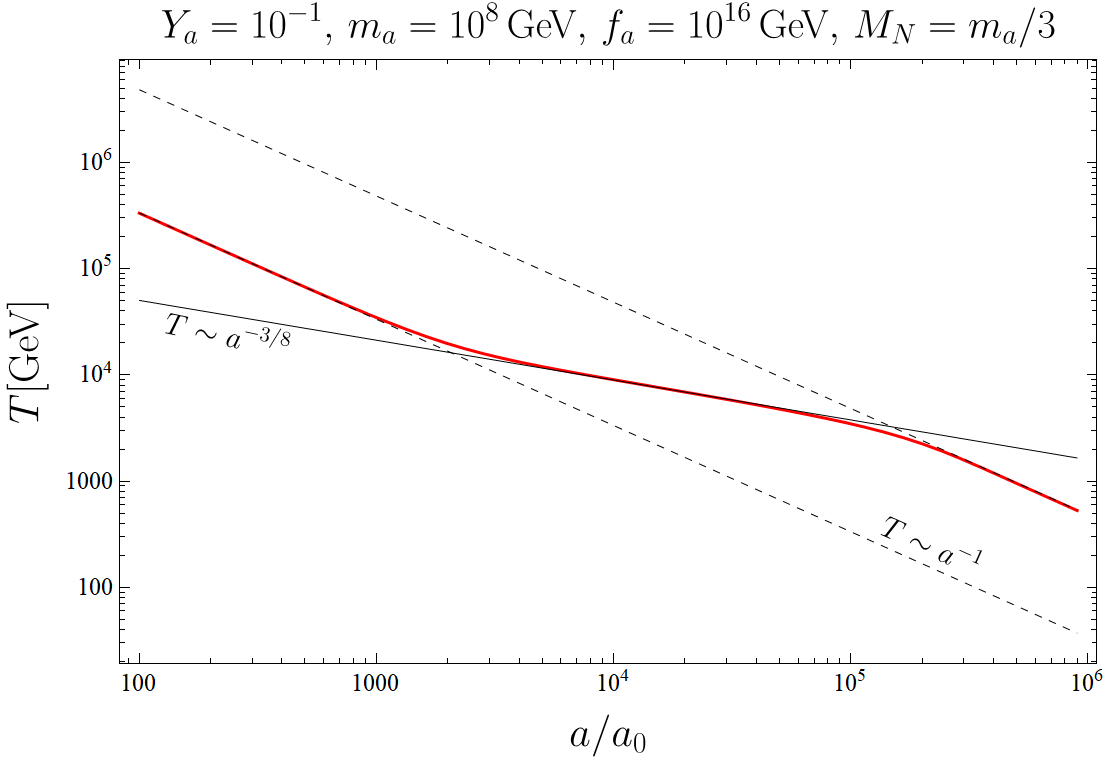}}
\hspace{2mm}
\subfigure
{\includegraphics[width=8.7cm]{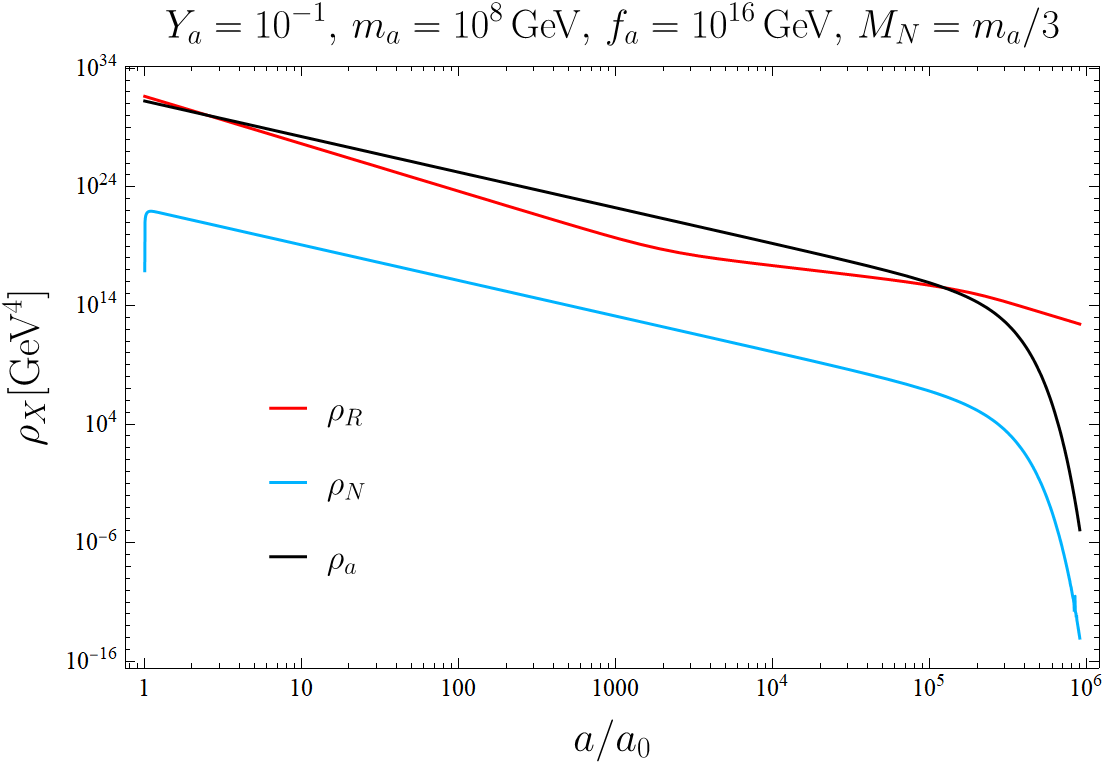}}
\caption{\small{Left plot: bath temperature $T$ as a function of the normalized scale factor $a/a_0$. We used some benchmark values for $Y_a$, $m_a$, $f_a$ to better highlight the features of the ALP-dominated epoch. In the ALP matter-dominated era, $T \sim a^{-3/8}$, while during radiation-dominated era as $T \sim a^{-1}$. Right plot: energy densities of the species involved.}}
\label{fig:temp_rho}
\end{figure}

\begin{figure}[h!]
\centering
\subfigure
{\includegraphics[width=8.7cm]{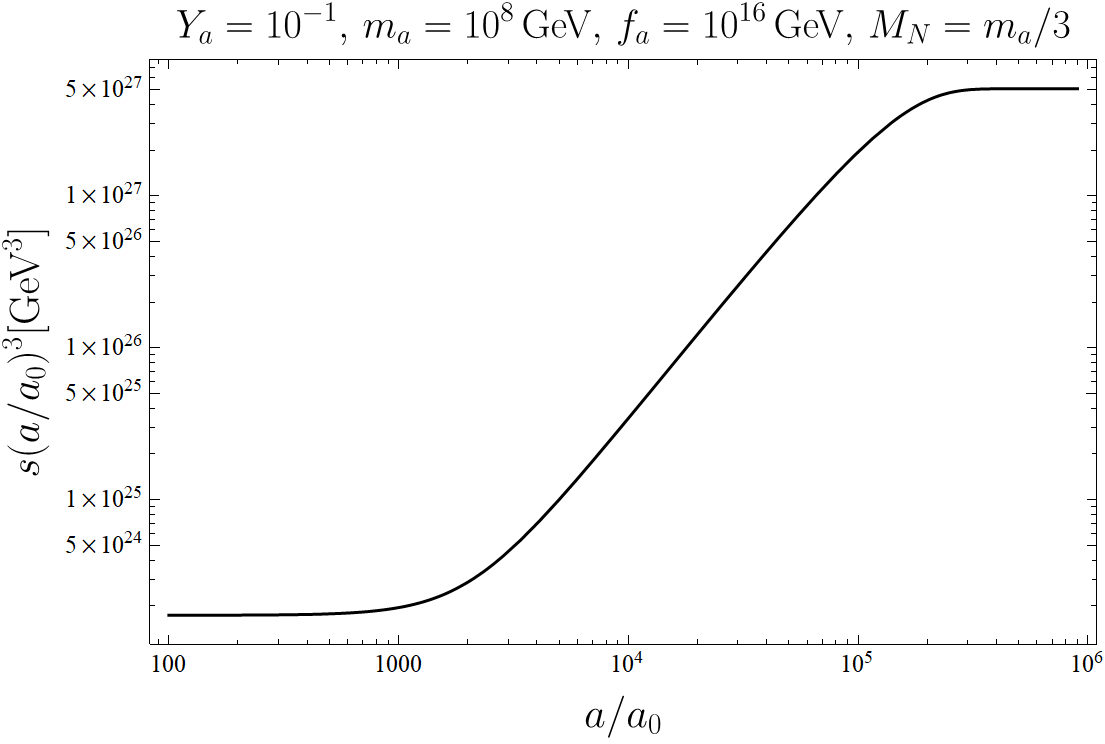}}
\hspace{2mm}
\subfigure
{\includegraphics[width=8.7cm]{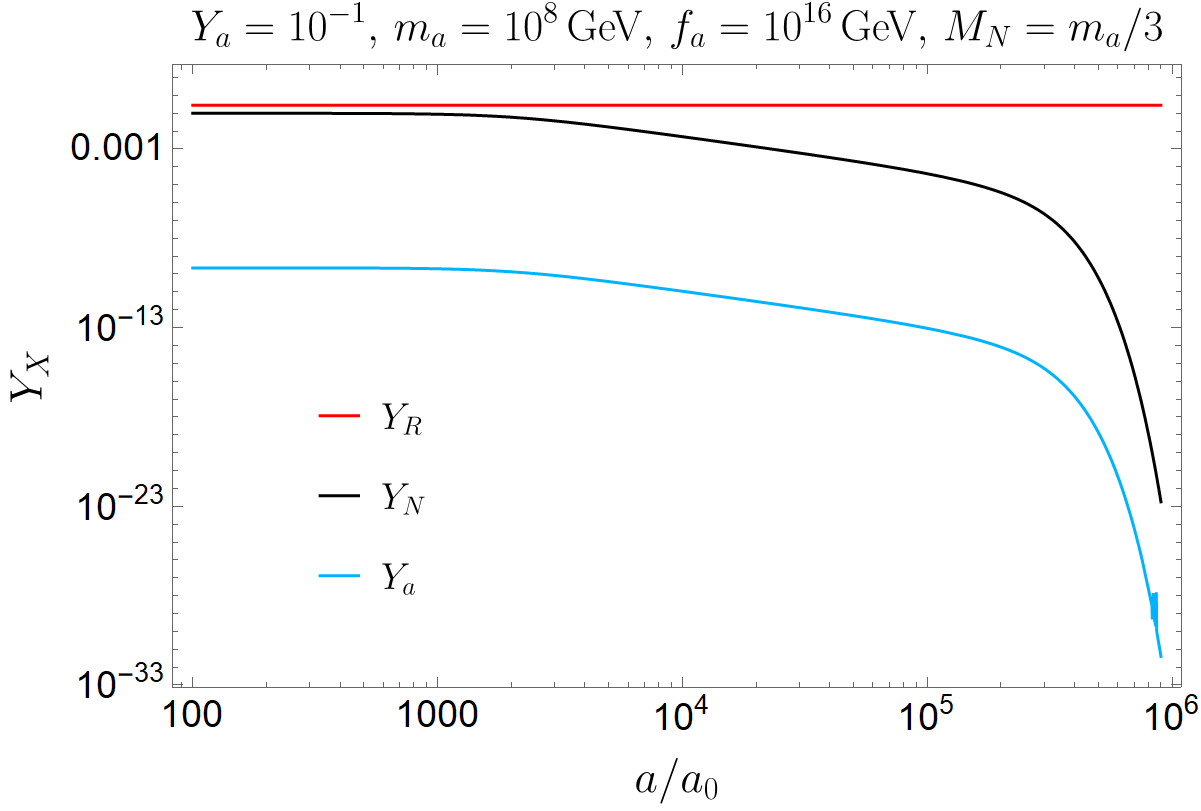}}
\caption{\small{Left plot: evolution of the total entropy $S= s (a/a_0)^3$. Right plot: Yields of species involved. Notice that the entropy injection in the thermal plasma, occurring when $10^3 \lesssim a/a_0 \lesssim 10^5$, dilutes both the RHN Yield and the ALP Yield.}}
\label{fig:entro_yield}
\end{figure}

\begin{figure}[h!]
\centering
\subfigure
{\includegraphics[width=9cm]{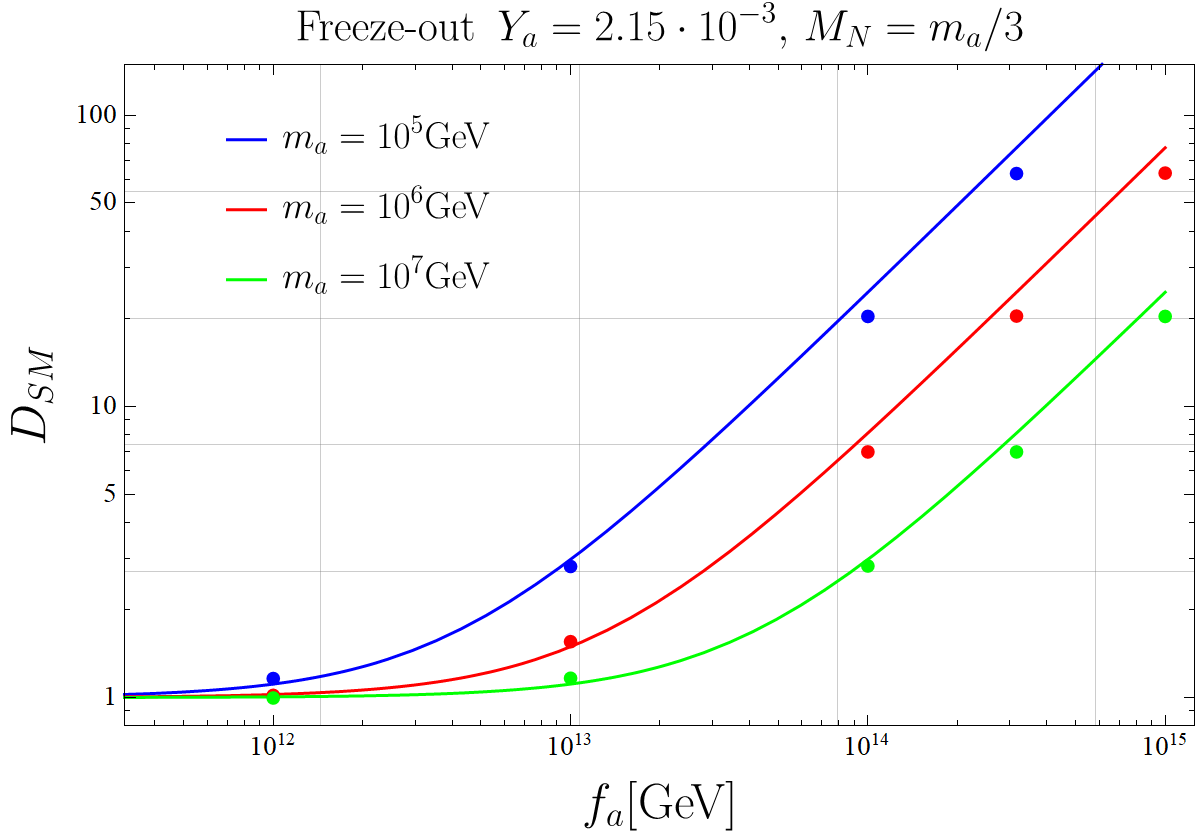}}
\caption{\small{We display the comparison of the analytical expression for the dilution factor, Eq.~(\ref{eq:DSMentropyratio}) (lines), with its numerical prediction computed from Eqs.\eqref{eq:dilsyst} (dots). We observe a maximal deviation of $15 \%$ when $D_{SM} \simeq 50$.}}
\label{fig:comparison}
\end{figure}

 In Eqs.\eqref{eq:dilsyst}, we can also plug the third relation into the fourth, the system of BEs becomes
 \bea 
 \label{eq:dilsystbis}
 \begin{cases}
    \dot{ \rho}_a = -3 \rho_{a}  - \frac{\Gamma_a}{H} \rho_a \\
    \dot{\rho}_N = -  3 \rho_N + \frac{\text{Br}_{N} \Gamma_a}{H}\rho_a - \frac{\Gamma_D}{H} \rho_N\\
  \dot{ Y}_{B-L}/ a_0=  - \frac{3}{H}Y_{B-L}\Big((\text{Br}_{g}+\text{Br}_{t}) \Gamma_a \rho_a +\Gamma_D \rho_N\Big) + \frac{\gamma_{D} \epsilon}{H s} \Big(\frac{Y_{N}}{Y_{N} ^{\rm eq}}-1\Big) - \frac{\gamma_D}{H s} \frac{Y_{B-L}}{2 Y_{l}^{\rm eq}}
    \end{cases}
 \, .
 \eea 
Solving Eqs.\eqref{eq:dilsyst}, we plot the evolution of the bath temperature (left panel) and energy densities $\rho_R$,  $\rho_N$,  $\rho_a$ (right panel) in Fig.~\ref{fig:temp_rho}, as a function of the normalized scale factor $a/a_0$. While the bath temperature scales as $T \sim a^{-1}$ during the radiation-dominated era, we find $T \sim a^{-3/8}$ during the ALP-induced matter-dominated epoch, consistently with \cite{PhysRevD.33.1585}. Moreover, we show the total entropy injection in the thermal plasma induced by the ALP decay in the left panel of Fig.~\ref{fig:entro_yield}. This leads to a dilution in both the Yields of RHN and ALP itself, as shown in the right panel of Fig.~\ref{fig:entro_yield}. Eventually, in Fig.~\ref{fig:comparison}, we compare the analytical expression of the dilution in Eq.\eqref{eq:DSMentropyratio} and the result from Eqs.\eqref{eq:dilsyst}. We observe that for dilution factor smaller than $D_{\rm SM} \lesssim 50$, the disagreement remains below 15 \%. In the main text, we will use the analytical expression for the dilution factor. 

 \clearpage 
 
\bibliographystyle{JHEP}
{\footnotesize
\bibliography{biblio}}

\providecommand{\href}[2]{#2}\begingroup\raggedright\begin{thebibliography}{10}

\bibitem{Planck:2015fie}
{\scshape Planck} collaboration, P.~A.~R. Ade et~al., \emph{{Planck 2015
  results. XIII. Cosmological parameters}},
  \href{http://dx.doi.org/10.1051/0004-6361/201525830}{\emph{Astron.
  Astrophys.} {\bf 594} (2016) A13},
  [\href{http://arxiv.org/abs/1502.01589}{{\tt 1502.01589}}].

\bibitem{Fields:2019pfx}
B.~D. Fields, K.~A. Olive, T.-H. Yeh and C.~Young, \emph{{Big-Bang
  Nucleosynthesis after Planck}},
  \href{http://dx.doi.org/10.1088/1475-7516/2020/03/010}{\emph{JCAP} {\bf 03}
  (2020) 010}, [\href{http://arxiv.org/abs/1912.01132}{{\tt 1912.01132}}].

\bibitem{Sakharov:1967dj}
A.~D. Sakharov, \emph{{Violation of CP Invariance, C asymmetry, and baryon
  asymmetry of the universe}},
  \href{http://dx.doi.org/10.1070/PU1991v034n05ABEH002497}{\emph{Pisma Zh.
  Eksp. Teor. Fiz.} {\bf 5} (1967) 32--35}.

\bibitem{Aoki:2006we}
Y.~Aoki, G.~Endrodi, Z.~Fodor, S.~D. Katz and K.~K. Szabo, \emph{{The Order of
  the quantum chromodynamics transition predicted by the standard model of
  particle physics}},
  \href{http://dx.doi.org/10.1038/nature05120}{\emph{Nature} {\bf 443} (2006)
  675--678}, [\href{http://arxiv.org/abs/hep-lat/0611014}{{\tt
  hep-lat/0611014}}].

\bibitem{Kajantie:1996mn}
K.~Kajantie, M.~Laine, K.~Rummukainen and M.~E. Shaposhnikov, \emph{{Is there
  a~ hot electroweak phase transition at $m_H \gtrsim m_W$?}},
  \href{http://dx.doi.org/10.1103/PhysRevLett.77.2887}{\emph{Phys. Rev. Lett.}
  {\bf 77} (1996) 2887--2890}, [\href{http://arxiv.org/abs/hep-ph/9605288}{{\tt
  hep-ph/9605288}}].

\bibitem{Gavela:1993ts}
M.~B. Gavela, P.~Hernandez, J.~Orloff and O.~Pene, \emph{{Standard model CP
  violation and baryon asymmetry}},
  \href{http://dx.doi.org/10.1142/S0217732394000629}{\emph{Mod. Phys. Lett. A}
  {\bf 9} (1994) 795--810}, [\href{http://arxiv.org/abs/hep-ph/9312215}{{\tt
  hep-ph/9312215}}].

\bibitem{Riotto:1998bt}
A.~Riotto, \emph{{Theories of baryogenesis}},  in \emph{{ICTP Summer School in
  High-Energy Physics and Cosmology}}, pp.~326--436, 7, 1998.
\newblock \href{http://arxiv.org/abs/hep-ph/9807454}{{\tt hep-ph/9807454}}.

\bibitem{Buchmuller:2021}
D.~Bodeker and W.~Buchmuller, \emph{{Baryogenesis from the weak scale to the
  grand unification scale}},
  \href{http://dx.doi.org/10.1103/RevModPhys.93.035004}{\emph{Rev.Mod.Phys.}
  {\bf 93, 3, 035004} (2021) 22}, [\href{http://arxiv.org/abs/2009.07294}{{\tt
  2009.07294}}].

\bibitem{FUKUGITA198645}
M.~Fukugita and T.~Yanagida, \emph{Barygenesis without grand unification},
  \href{http://dx.doi.org/https://doi.org/10.1016/0370-2693(86)91126-3}{\emph{Physics
  Letters B} {\bf 174} (1986) 45--47}.

\bibitem{Davidson:2008bu}
S.~Davidson, E.~Nardi and Y.~Nir, \emph{{Leptogenesis}},
  \href{http://dx.doi.org/10.1016/j.physrep.2008.06.002}{\emph{Phys. Rept.}
  {\bf 466} (2008) 105--177}, [\href{http://arxiv.org/abs/0802.2962}{{\tt
  0802.2962}}].

\bibitem{Buchm_ller_2005}
W.~Buchmüller, P.~D. Bari and M.~Plümacher, \emph{Leptogenesis for
  pedestrians}, \href{http://dx.doi.org/10.1016/j.aop.2004.02.003}{\emph{Annals
  of Physics} {\bf 315} (feb, 2005) 305--351}.

\bibitem{Chung:1998rq}
D.~J.~H. Chung, E.~W. Kolb and A.~Riotto, \emph{{Production of massive
  particles during reheating}},
  \href{http://dx.doi.org/10.1103/PhysRevD.60.063504}{\emph{Phys. Rev. D} {\bf
  60} (1999) 063504}, [\href{http://arxiv.org/abs/hep-ph/9809453}{{\tt
  hep-ph/9809453}}].

\bibitem{Giudice:1999fb}
G.~F. Giudice, M.~Peloso, A.~Riotto and I.~Tkachev, \emph{{Production of
  massive fermions at preheating and leptogenesis}},
  \href{http://dx.doi.org/10.1088/1126-6708/1999/08/014}{\emph{JHEP} {\bf 08}
  (1999) 014}, [\href{http://arxiv.org/abs/hep-ph/9905242}{{\tt
  hep-ph/9905242}}].

\bibitem{Hahn-Woernle:2008tsk}
F.~Hahn-Woernle and M.~Plumacher, \emph{{Effects of reheating on
  leptogenesis}},
  \href{http://dx.doi.org/10.1016/j.nuclphysb.2008.07.032}{\emph{Nucl. Phys. B}
  {\bf 806} (2009) 68--83}, [\href{http://arxiv.org/abs/0801.3972}{{\tt
  0801.3972}}].

\bibitem{Zhang:2023oyo}
X.~Zhang, \emph{{Towards a systematic study of non-thermal leptogenesis from
  inflaton decays}},  \href{http://arxiv.org/abs/2311.05824}{{\tt 2311.05824}}.

\bibitem{Katz:2016adq}
A.~Katz and A.~Riotto, \emph{{Baryogenesis and Gravitational Waves from Runaway
  Bubble Collisions}},
  \href{http://dx.doi.org/10.1088/1475-7516/2016/11/011}{\emph{JCAP} {\bf 11}
  (2016) 011}, [\href{http://arxiv.org/abs/1608.00583}{{\tt 1608.00583}}].

\bibitem{Azatov:2021irb}
A.~Azatov, M.~Vanvlasselaer and W.~Yin, \emph{{Baryogenesis via relativistic
  bubble walls}}, \href{http://dx.doi.org/10.1007/JHEP10(2021)043}{\emph{JHEP}
  {\bf 10} (2021) 043}, [\href{http://arxiv.org/abs/2106.14913}{{\tt
  2106.14913}}].

\bibitem{Huang:2022vkf}
P.~Huang and K.-P. Xie, \emph{{Leptogenesis triggered by a first-order phase
  transition}}, \href{http://dx.doi.org/10.1007/JHEP09(2022)052}{\emph{JHEP}
  {\bf 09} (2022) 052}, [\href{http://arxiv.org/abs/2206.04691}{{\tt
  2206.04691}}].

\bibitem{Chun:2023ezg}
E.~J. Chun, T.~P. Dutka, T.~H. Jung, X.~Nagels and M.~Vanvlasselaer,
  \emph{{Bubble-assisted Leptogenesis}},
  \href{http://arxiv.org/abs/2305.10759}{{\tt 2305.10759}}.

\bibitem{Dichtl:2023xqd}
M.~Dichtl, J.~Nava, S.~Pascoli and F.~Sala, \emph{{Baryogenesis and
  leptogenesis from supercooled confinement}},
  \href{http://dx.doi.org/10.1007/JHEP02(2024)059}{\emph{JHEP} {\bf 02} (2024)
  059}, [\href{http://arxiv.org/abs/2312.09282}{{\tt 2312.09282}}].

\bibitem{Daido:2015gqa}
R.~Daido, N.~Kitajima and F.~Takahashi, \emph{{Axion domain wall
  baryogenesis}},
  \href{http://dx.doi.org/10.1088/1475-7516/2015/07/046}{\emph{JCAP} {\bf 07}
  (2015) 046}, [\href{http://arxiv.org/abs/1504.07917}{{\tt 1504.07917}}].

\bibitem{Chun:2023eqc}
E.~J. Chun and T.~H. Jung, \emph{{Leptogenesis driven by a Majoron}},
  \href{http://dx.doi.org/10.1103/PhysRevD.109.095004}{\emph{Phys. Rev. D} {\bf
  109} (2024) 095004}, [\href{http://arxiv.org/abs/2311.09005}{{\tt
  2311.09005}}].

\bibitem{CidVidal:2018blh}
X.~Cid~Vidal, A.~Mariotti, D.~Redigolo, F.~Sala and K.~Tobioka, \emph{{New
  Axion Searches at Flavor Factories}},
  \href{http://dx.doi.org/10.1007/JHEP01(2019)113}{\emph{JHEP} {\bf 01} (2019)
  113}, [\href{http://arxiv.org/abs/1810.09452}{{\tt 1810.09452}}].

\bibitem{Weinberg:1977ma}
S.~Weinberg, \emph{{A New Light Boson?}},
  \href{http://dx.doi.org/10.1103/PhysRevLett.40.223}{\emph{Phys. Rev. Lett.}
  {\bf 40} (1978) 223--226}.

\bibitem{Wilczek:1977pj}
F.~Wilczek, \emph{{Problem of Strong $P$ and $T$ Invariance in the Presence of
  Instantons}}, \href{http://dx.doi.org/10.1103/PhysRevLett.40.279}{\emph{Phys.
  Rev. Lett.} {\bf 40} (1978) 279--282}.

\bibitem{Kamionkowski:1992mf}
M.~Kamionkowski and J.~March-Russell, \emph{{Planck scale physics and the
  Peccei-Quinn mechanism}},
  \href{http://dx.doi.org/10.1016/0370-2693(92)90492-M}{\emph{Phys. Lett. B}
  {\bf 282} (1992) 137--141}, [\href{http://arxiv.org/abs/hep-th/9202003}{{\tt
  hep-th/9202003}}].

\bibitem{Holman:1992us}
R.~Holman, S.~D.~H. Hsu, T.~W. Kephart, E.~W. Kolb, R.~Watkins and L.~M.
  Widrow, \emph{{Solutions to the strong CP problem in a world with gravity}},
  \href{http://dx.doi.org/10.1016/0370-2693(92)90491-L}{\emph{Phys. Lett. B}
  {\bf 282} (1992) 132--136}, [\href{http://arxiv.org/abs/hep-ph/9203206}{{\tt
  hep-ph/9203206}}].

\bibitem{Ghigna:1992iv}
S.~Ghigna, M.~Lusignoli and M.~Roncadelli, \emph{{Instability of the invisible
  axion}}, \href{http://dx.doi.org/10.1016/0370-2693(92)90019-Z}{\emph{Phys.
  Lett. B} {\bf 283} (1992) 278--281}.

\bibitem{Barr:1992qq}
S.~M. Barr and D.~Seckel, \emph{{Planck scale corrections to axion models}},
  \href{http://dx.doi.org/10.1103/PhysRevD.46.539}{\emph{Phys. Rev. D} {\bf 46}
  (1992) 539--549}.

\bibitem{Nelson:1993nf}
A.~E. Nelson and N.~Seiberg, \emph{{R symmetry breaking versus supersymmetry
  breaking}}, \href{http://dx.doi.org/10.1016/0550-3213(94)90577-0}{\emph{Nucl.
  Phys. B} {\bf 416} (1994) 46--62},
  [\href{http://arxiv.org/abs/hep-ph/9309299}{{\tt hep-ph/9309299}}].

\bibitem{Intriligator:2007py}
K.~A. Intriligator, N.~Seiberg and D.~Shih, \emph{{Supersymmetry breaking,
  R-symmetry breaking and metastable vacua}},
  \href{http://dx.doi.org/10.1088/1126-6708/2007/07/017}{\emph{JHEP} {\bf 07}
  (2007) 017}, [\href{http://arxiv.org/abs/hep-th/0703281}{{\tt
  hep-th/0703281}}].

\bibitem{Bellazzini:2017neg}
B.~Bellazzini, A.~Mariotti, D.~Redigolo, F.~Sala and J.~Serra, \emph{{R-axion
  at colliders}},
  \href{http://dx.doi.org/10.1103/PhysRevLett.119.141804}{\emph{Phys. Rev.
  Lett.} {\bf 119} (2017) 141804}, [\href{http://arxiv.org/abs/1702.02152}{{\tt
  1702.02152}}].

\bibitem{Martin:1997ns}
S.~P. Martin, \emph{{A Supersymmetry primer}},
  \href{http://dx.doi.org/10.1142/9789812839657_0001}{\emph{Adv. Ser. Direct.
  High Energy Phys.} {\bf 18} (1998) 1--98},
  [\href{http://arxiv.org/abs/hep-ph/9709356}{{\tt hep-ph/9709356}}].

\bibitem{Terning:2006bq}
J.~Terning, \emph{{Modern supersymmetry: Dynamics and duality}}.
\newblock 2006,
  \href{http://dx.doi.org/10.1093/acprof:oso/9780198567639.001.0001}{10.1093/acprof:oso/9780198567639.001.0001}.

\bibitem{Peccei:1977hh}
R.~D. Peccei and H.~R. Quinn, \emph{{CP Conservation in the Presence of
  Instantons}},
  \href{http://dx.doi.org/10.1103/PhysRevLett.38.1440}{\emph{Phys. Rev. Lett.}
  {\bf 38} (1977) 1440--1443}.

\bibitem{Peccei:1977ur}
R.~D. Peccei and H.~R. Quinn, \emph{{Constraints Imposed by CP Conservation in
  the Presence of Instantons}},
  \href{http://dx.doi.org/10.1103/PhysRevD.16.1791}{\emph{Phys. Rev. D} {\bf
  16} (1977) 1791--1797}.

\bibitem{Zhitnitsky:1980tq}
A.~R. Zhitnitsky, \emph{{On Possible Suppression of the Axion Hadron
  Interactions. (In Russian)}}, {\emph{Sov. J. Nucl. Phys.} {\bf 31} (1980)
  260}.

\bibitem{Dine:1981rt}
M.~Dine, W.~Fischler and M.~Srednicki, \emph{{A Simple Solution to the Strong
  CP Problem with a Harmless Axion}},
  \href{http://dx.doi.org/10.1016/0370-2693(81)90590-6}{\emph{Phys. Lett. B}
  {\bf 104} (1981) 199--202}.

\bibitem{Dine:2009sw}
M.~Dine, G.~Festuccia and Z.~Komargodski, \emph{{A Bound on the
  Superpotential}},
  \href{http://dx.doi.org/10.1007/JHEP03(2010)011}{\emph{JHEP} {\bf 03} (2010)
  011}, [\href{http://arxiv.org/abs/0910.2527}{{\tt 0910.2527}}].

\bibitem{Bellazzini:2016xrt}
B.~Bellazzini, \emph{{Softness and amplitudes\textquoteright{} positivity for
  spinning particles}},
  \href{http://dx.doi.org/10.1007/JHEP02(2017)034}{\emph{JHEP} {\bf 02} (2017)
  034}, [\href{http://arxiv.org/abs/1605.06111}{{\tt 1605.06111}}].

\bibitem{Salvio_2014}
A.~Salvio, A.~Strumia and W.~Xue, \emph{Thermal axion production},
  \href{http://dx.doi.org/10.1088/1475-7516/2014/01/011}{\emph{Journal of
  Cosmology and Astroparticle Physics} {\bf 2014} (jan, 2014) 011--011}.

\bibitem{Bouzoud:2024bom}
K.~Bouzoud and J.~Ghiglieri, \emph{{Thermal axion production at hard and soft
  momenta}},  \href{http://arxiv.org/abs/2404.06113}{{\tt 2404.06113}}.

\bibitem{Hamada_2014}
Y.~Hamada, K.~Kamada, T.~Kobayashi and Y.~Ookouchi, \emph{More on cosmological
  constraints on spontaneous r-symmetry breaking models},
  \href{http://dx.doi.org/10.1088/1475-7516/2014/01/024}{\emph{Journal of
  Cosmology and Astroparticle Physics} {\bf 2014} (jan, 2014) 024--024}.

\bibitem{Minkowski:1977sc}
P.~Minkowski, \emph{{$\mu \to e\gamma$ at a Rate of One Out of $10^{9}$ Muon
  Decays?}}, \href{http://dx.doi.org/10.1016/0370-2693(77)90435-X}{\emph{Phys.
  Lett. B} {\bf 67} (1977) 421--428}.

\bibitem{Yanagida:1979as}
T.~Yanagida, \emph{{Horizontal gauge symmetry and masses of neutrinos}},
  {\emph{Conf. Proc. C} {\bf 7902131} (1979) 95--99}.

\bibitem{Mohapatra:1979ia}
R.~N. Mohapatra and G.~Senjanovic, \emph{{Neutrino Mass and Spontaneous Parity
  Nonconservation}},
  \href{http://dx.doi.org/10.1103/PhysRevLett.44.912}{\emph{Phys. Rev. Lett.}
  {\bf 44} (1980) 912}.

\bibitem{Schechter:1980gr}
J.~Schechter and J.~W.~F. Valle, \emph{{Neutrino Masses in SU(2) x U(1)
  Theories}}, \href{http://dx.doi.org/10.1103/PhysRevD.22.2227}{\emph{Phys.
  Rev. D} {\bf 22} (1980) 2227}.

\bibitem{Plümacher1997549}
M.~Plümacher, \emph{Baryogenesis and lepton number violation},
  \href{http://dx.doi.org/10.1007/s002880050418}{\emph{Zeitschrift fur Physik
  C-Particles and Fields} {\bf 74} (1997) 549 – 559}.

\bibitem{Gonzalez_Garcia_2021}
M.~C. Gonzalez-Garcia, M.~Maltoni and T.~Schwetz, \emph{{NuFIT}: Three-flavour
  global analyses of neutrino oscillation experiments},
  \href{http://dx.doi.org/10.3390/universe7120459}{\emph{Universe} {\bf 7}
  (nov, 2021) 459}.

\bibitem{Moffat:2018wke}
K.~Moffat, S.~Pascoli, S.~T. Petcov, H.~Schulz and J.~Turner,
  \emph{{Three-flavored nonresonant leptogenesis at intermediate scales}},
  \href{http://dx.doi.org/10.1103/PhysRevD.98.015036}{\emph{Phys. Rev. D} {\bf
  98} (2018) 015036}, [\href{http://arxiv.org/abs/1804.05066}{{\tt
  1804.05066}}].

\bibitem{Granelli:2021fyc}
A.~Granelli, K.~Moffat and S.~T. Petcov, \emph{{Aspects of high scale
  leptogenesis with low-energy leptonic CP violation}},
  \href{http://dx.doi.org/10.1007/JHEP11(2021)149}{\emph{JHEP} {\bf 11} (2021)
  149}, [\href{http://arxiv.org/abs/2107.02079}{{\tt 2107.02079}}].

\bibitem{Davidson_2002}
S.~Davidson and A.~Ibarra, \emph{A lower bound on the right-handed neutrino
  mass from leptogenesis},
  \href{http://dx.doi.org/10.1016/s0370-2693(02)01735-5}{\emph{Physics Letters
  B} {\bf 535} (may, 2002) 25--32}.

\bibitem{Hamaguchi_2002}
K.~Hamaguchi, H.~Murayama and T.~Yanagida, \emph{Leptogenesis from a
  greater-dominated early universe},
  \href{http://dx.doi.org/10.1103/physrevd.65.043512}{\emph{Physical Review D}
  {\bf 65} (jan, 2002) }.

\bibitem{Pilaftsis:2003gt}
A.~Pilaftsis and T.~E.~J. Underwood, \emph{{Resonant leptogenesis}},
  \href{http://dx.doi.org/10.1016/j.nuclphysb.2004.05.029}{\emph{Nucl. Phys. B}
  {\bf 692} (2004) 303--345}, [\href{http://arxiv.org/abs/hep-ph/0309342}{{\tt
  hep-ph/0309342}}].

\bibitem{Pilaftsis_2005}
A.~Pilaftsis and T.~E.~J. Underwood, \emph{Electroweak-scale resonant
  leptogenesis},
  \href{http://dx.doi.org/10.1103/physrevd.72.113001}{\emph{Physical Review D}
  {\bf 72} (dec, 2005) }.

\bibitem{Granelli:2020ysj}
A.~Granelli, K.~Moffat and S.~T. Petcov, \emph{{Flavoured resonant leptogenesis
  at sub-TeV scales}},
  \href{http://dx.doi.org/10.1016/j.nuclphysb.2021.115597}{\emph{Nucl. Phys. B}
  {\bf 973} (2021) 115597}, [\href{http://arxiv.org/abs/2009.03166}{{\tt
  2009.03166}}].

\bibitem{Pilaftsis_2004}
A.~Pilaftsis and T.~E. Underwood, \emph{Resonant leptogenesis},
  \href{http://dx.doi.org/10.1016/j.nuclphysb.2004.05.029}{\emph{Nuclear
  Physics B} {\bf 692} (aug, 2004) 303--345}.

\bibitem{Nardi:2007jp}
E.~Nardi, J.~Racker and E.~Roulet, \emph{{CP violation in scatterings, three
  body processes and the Boltzmann equations for leptogenesis}},
  \href{http://dx.doi.org/10.1088/1126-6708/2007/09/090}{\emph{JHEP} {\bf 09}
  (2007) 090}, [\href{http://arxiv.org/abs/0707.0378}{{\tt 0707.0378}}].

\bibitem{Nemevsek:2022anh}
M.~Nemev\v{s}ek and Y.~Zhang, \emph{{Dark Matter Dilution Mechanism through the
  Lens of Large-Scale Structure}},
  \href{http://dx.doi.org/10.1103/PhysRevLett.130.121002}{\emph{Phys. Rev.
  Lett.} {\bf 130} (2023) 121002}, [\href{http://arxiv.org/abs/2206.11293}{{\tt
  2206.11293}}].

\bibitem{Cirelli_2019}
M.~Cirelli, Y.~Gouttenoire, K.~Petraki and F.~Sala, \emph{Homeopathic dark
  matter, or how diluted heavy substances produce high energy cosmic rays},
  \href{http://dx.doi.org/10.1088/1475-7516/2019/02/014}{\emph{Journal of
  Cosmology and Astroparticle Physics} {\bf 2019} (feb, 2019) 014--014}.

\bibitem{Hahn-Woernle:2009jyb}
F.~Hahn-Woernle, M.~Plumacher and Y.~Y.~Y. Wong, \emph{{Full Boltzmann
  equations for leptogenesis including scattering}},
  \href{http://dx.doi.org/10.1088/1475-7516/2009/08/028}{\emph{JCAP} {\bf 08}
  (2009) 028}, [\href{http://arxiv.org/abs/0907.0205}{{\tt 0907.0205}}].

\bibitem{D_Eramo_2021}
F.~D’Eramo and A.~Lenoci, \emph{Lower mass bounds on fimp dark matter
  produced via freeze-in},
  \href{http://dx.doi.org/10.1088/1475-7516/2021/10/045}{\emph{Journal of
  Cosmology and Astroparticle Physics} {\bf 2021} (Oct., 2021) 045}.

\bibitem{Pilaftsis:1997jf}
A.~Pilaftsis, \emph{{CP violation and baryogenesis due to heavy Majorana
  neutrinos}}, \href{http://dx.doi.org/10.1103/PhysRevD.56.5431}{\emph{Phys.
  Rev. D} {\bf 56} (1997) 5431--5451},
  [\href{http://arxiv.org/abs/hep-ph/9707235}{{\tt hep-ph/9707235}}].

\bibitem{Giudice:1998ic}
G.~F. Giudice and R.~Rattazzi, \emph{{Gauge-mediated supersymmetry breaking}},
  \href{http://dx.doi.org/10.1142/9789812839657_0015}{\emph{Adv. Ser. Direct.
  High Energy Phys.} {\bf 18} (1998) 355--377}.

\bibitem{Rychkov:2007uq}
V.~S. Rychkov and A.~Strumia, \emph{{Thermal production of gravitinos}},
  \href{http://dx.doi.org/10.1103/PhysRevD.75.075011}{\emph{Phys. Rev. D} {\bf
  75} (2007) 075011}, [\href{http://arxiv.org/abs/hep-ph/0701104}{{\tt
  hep-ph/0701104}}].

\bibitem{Moroi:1993mb}
T.~Moroi, H.~Murayama and M.~Yamaguchi, \emph{{Cosmological constraints on the
  light stable gravitino}},
  \href{http://dx.doi.org/10.1016/0370-2693(93)91434-O}{\emph{Phys. Lett. B}
  {\bf 303} (1993) 289--294}.

\bibitem{Kawasaki:1994af}
M.~Kawasaki and T.~Moroi, \emph{{Gravitino production in the inflationary
  universe and the effects on big bang nucleosynthesis}},
  \href{http://dx.doi.org/10.1143/PTP.93.879}{\emph{Prog. Theor. Phys.} {\bf
  93} (1995) 879--900}, [\href{http://arxiv.org/abs/hep-ph/9403364}{{\tt
  hep-ph/9403364}}].

\bibitem{Kawasaki:2008qe}
M.~Kawasaki, K.~Kohri, T.~Moroi and A.~Yotsuyanagi, \emph{{Big-Bang
  Nucleosynthesis and Gravitino}},
  \href{http://dx.doi.org/10.1103/PhysRevD.78.065011}{\emph{Phys. Rev. D} {\bf
  78} (2008) 065011}, [\href{http://arxiv.org/abs/0804.3745}{{\tt 0804.3745}}].

\bibitem{Brignole:1997sk}
A.~Brignole, F.~Feruglio and F.~Zwirner, \emph{{Signals of a superlight
  gravitino at $e^+ e^-$ colliders when the other superparticles are heavy}},
  \href{http://dx.doi.org/10.1016/S0550-3213(97)00825-0}{\emph{Nucl. Phys. B}
  {\bf 516} (1998) 13--28}, [\href{http://arxiv.org/abs/hep-ph/9711516}{{\tt
  hep-ph/9711516}}].

\bibitem{Planck:2018vyg}
{\scshape Planck} collaboration, N.~Aghanim et~al., \emph{{Planck 2018 results.
  VI. Cosmological parameters}},
  \href{http://dx.doi.org/10.1051/0004-6361/201833910}{\emph{Astron.
  Astrophys.} {\bf 641} (2020) A6},
  [\href{http://arxiv.org/abs/1807.06209}{{\tt 1807.06209}}].

\bibitem{Mohapatra:1986aw}
R.~N. Mohapatra, \emph{{Mechanism for Understanding Small Neutrino Mass in
  Superstring Theories}},
  \href{http://dx.doi.org/10.1103/PhysRevLett.56.561}{\emph{Phys. Rev. Lett.}
  {\bf 56} (1986) 561--563}.

\bibitem{Mohapatra:1986bd}
R.~N. Mohapatra and J.~W.~F. Valle, \emph{{Neutrino Mass and Baryon Number
  Nonconservation in Superstring Models}},
  \href{http://dx.doi.org/10.1103/PhysRevD.34.1642}{\emph{Phys. Rev. D} {\bf
  34} (1986) 1642}.

\bibitem{Li:2023tbx}
P.~Li, Z.~Liu and K.-F. Lyu, \emph{{Heavy neutral leptons at muon colliders}},
  \href{http://dx.doi.org/10.1007/JHEP03(2023)231}{\emph{JHEP} {\bf 03} (2023)
  231}, [\href{http://arxiv.org/abs/2301.07117}{{\tt 2301.07117}}].

\bibitem{Chikashige:1980qk}
Y.~Chikashige, R.~N. Mohapatra and R.~D. Peccei, \emph{{Spontaneously Broken
  Lepton Number and Cosmological Constraints on the Neutrino Mass Spectrum}},
  \href{http://dx.doi.org/10.1103/PhysRevLett.45.1926}{\emph{Phys. Rev. Lett.}
  {\bf 45} (1980) 1926}.

\bibitem{Chikashige:1980ui}
Y.~Chikashige, R.~N. Mohapatra and R.~D. Peccei, \emph{{Are There Real
  Goldstone Bosons Associated with Broken Lepton Number?}},
  \href{http://dx.doi.org/10.1016/0370-2693(81)90011-3}{\emph{Phys. Lett. B}
  {\bf 98} (1981) 265--268}.

\bibitem{Gelmini:1980re}
G.~B. Gelmini and M.~Roncadelli, \emph{{Left-Handed Neutrino Mass Scale and
  Spontaneously Broken Lepton Number}},
  \href{http://dx.doi.org/10.1016/0370-2693(81)90559-1}{\emph{Phys. Lett. B}
  {\bf 99} (1981) 411--415}.

\bibitem{Tong:2024lmi}
J.-Y. Tong, Z.-H. Yu and H.-H. Zhang, \emph{{Leptogenesis assisted by scalar
  decays}},  \href{http://arxiv.org/abs/2406.13468}{{\tt 2406.13468}}.

\bibitem{Babu:2009pi}
K.~S. Babu, Y.~Meng and Z.~Tavartkiladze, \emph{{New Ways to Leptogenesis with
  Gauged B-L Symmetry}},
  \href{http://dx.doi.org/10.1016/j.physletb.2009.09.036}{\emph{Phys. Lett. B}
  {\bf 681} (2009) 37--43}, [\href{http://arxiv.org/abs/0901.1044}{{\tt
  0901.1044}}].

\bibitem{Komargodski:2009rz}
Z.~Komargodski and N.~Seiberg, \emph{{From Linear SUSY to Constrained
  Superfields}},
  \href{http://dx.doi.org/10.1088/1126-6708/2009/09/066}{\emph{JHEP} {\bf 09}
  (2009) 066}, [\href{http://arxiv.org/abs/0907.2441}{{\tt 0907.2441}}].

\bibitem{Gorbatov:2008qa}
E.~Gorbatov and M.~Sudano, \emph{{Sparticle Masses in Higgsed Gauge
  Mediation}},
  \href{http://dx.doi.org/10.1088/1126-6708/2008/10/066}{\emph{JHEP} {\bf 10}
  (2008) 066}, [\href{http://arxiv.org/abs/0802.0555}{{\tt 0802.0555}}].

\bibitem{Giudice_2004}
G.~Giudice, A.~Notari, M.~Raidal, A.~Riotto and A.~Strumia, \emph{Towards a
  complete theory of thermal leptogenesis in the sm and mssm},
  \href{http://dx.doi.org/10.1016/j.nuclphysb.2004.02.019}{\emph{Nuclear
  Physics B} {\bf 685} (May, 2004) 89–149}.

\bibitem{PhysRevD.33.1585}
R.~J. Scherrer and M.~S. Turner, \emph{On the relic, cosmic abundance of
  stable, weakly interacting massive particles},
  \href{http://dx.doi.org/10.1103/PhysRevD.33.1585}{\emph{Phys. Rev. D} {\bf
  33} (Mar, 1986) 1585--1589}.

\end{thebibliography}\endgroup

\end{document}